# Thermodynamics of mixtures with strongly negative deviations from Raoult's law. XVII. Permittivities and refractive indices for alkan-1-ol + *N,N*-diethylethanamine systems at (293.15-303.15) K. Application of the Kirkwood-Fröhlich model

Fernando Hevia, Juan Antonio González*, Ana Cobos, Isaías García de la Fuente, L.F. Sanz

G.E.T.E.F., Departamento de Física Aplicada, Facultad de Ciencias, Universidad de Valladolid. Paseo de Belén, 7, 47011 Valladolid, Spain.

*e-mail: jagl@termo.uva.es; Tel: +34 983 423757

# Abstract

Relative permittivities at 1 MHz, $\varepsilon_r$, at 0.1 MPa and (293.15-303.15) K and refractive indices, $n_D$, at similar conditions have been measured for the alkan-1-ol (methanol, propan-1-ol, butan-1-ol, pentan-1-ol or heptan-1-ol) + *N,N*-diethylethanamine (TEA) systems. Positive values of the excess permittivities, $\varepsilon_r^E$, are encountered for the methanol system at high alcohol concentrations. The remaining mixtures are characterized by negative $\varepsilon_r^E$ values over the whole composition range. At $\phi_1$ (volume fraction) = 0.5, $\varepsilon_r^E$ changes in the order: methanol > propan-1-ol > butan-1-ol < pentan-1-ol < heptan-1-ol. Mixtures formed by alkan-1-ol and an isomeric amine, hexan-1-amine (HxA) or *N*-propylpropan-1-amine (DPA) or cyclohexylamine, behave similarly. This has been explained in terms of the lower and weaker self-association of longer alkan-1-ols. From the permittivity data, it is shown that: (i) (alkan-1-ol)-TEA interactions contribute positively to $\varepsilon_r^E$; (ii) TEA is an effective breaker of the network of the alkan-1-ols; (iii) structural effects, which are very important for the volumetric and calorimetric data of alkan-1-ol + TEA systems, are also relevant when evaluating dielectric data. This is confirmed by the comparison of $\varepsilon_r^E$ measurements for alkan-1-ol + aliphatic amine mixtures; (iv) the aromaticity effect (i.e., the replacement of TEA by pyridine in systems with a given alkan-1-ol) leads to an increase of the mixture polarization. Calculations conducted in the framework of the Kirkwood-Fröhlich model are consistent with the previous statements.

# 1. Introduction

Alkan-1-ol + linear primary or secondary amine mixtures are characterized by showing strongly negative deviations from Raoult's law [1]. As a consequence, their excess molar Gibbs energies, $G_m^E$, and enthalpies, $H_m^E$, are both negative, the former even at rather high temperatures. For example, for the methanol + butan-1-amine mixture at equimolar composition, $G_m^E = -799$ J·mol$^{-1}$ ($T = 348.15$ K) [2] and $H_m^E = -3767$ J.mol$^{-1}$ ($T = 298.15$ K) [3]. On the other hand, this type of solutions is characterized by large solvation effects. In fact, the equilibrium constants, $K_{AB}$, related to the formation of linear chains of the type $A_n(\text{alkan-1-ol}) + B_m(\text{linear amine}) € A_nB_m$, calculated by means of the ERAS model [1,4], are rather large and the corresponding enthalpies of hydrogen bonds between alkan-1-ol and amine, $\Delta h_{AB}^*$, are large and negative [1,5-8], and the same occurs for $\Delta v_{AB}^*$, the association volume of component A with B. Thus, for the methanol + hexan-1-amine (HxA) system, $K_{AB}$ = 2500 ($T$ = 298.15 K); $\Delta h_{AB}^*$ = –42.4 kJ·mol$^{-1}$; $\Delta v_{AB}^*$= 9.1 cm$^3$·mol$^{-1}$ [5]. Interestingly, $\Delta h_{AB}^*$ values are lower than those related to the H-bonds between alkan-1-ol molecules (–25.1 kJ·mol$^{-1}$ [1,3,4]). That is, (alkan-1-ol)-amine interactions are stronger than those between molecules of alkan-1-ol, which explains the large and negative $H_m^E$ values observed for these systems. Hereafter, we are referring, except when indicated, to excess molar functions at equimolar composition and 298.15 K.

Alkan-1-ol + *N,N*-diethylethanamine (TEA) mixtures are somewhat different. Some of their most relevant features are the following. (i) $G_m^E$ values are usually positive: 284 J·mol$^{-1}$ for the methanol-containing mixture ($T$ = 303.2 K) [9]; (ii) $H_m^E$ values are negative but lower in absolute value than for alkan-1-ol + linear primary or secondary amine systems: –1871 J·mol$^{-1}$ [10], and –1520 J·mol$^{-1}$ [11] for the solutions with methanol, or ethanol, respectively; (iii) Results on $V_m^E$, excess molar volume, are much more negative than for the systems with the isomeric amines HxA or *N*-propylpropan-1-amine (DPA). For example, $V_m^E$ (propan-1-ol) / 10$^{-6}$ m$^3$·mol$^{-1}$ = –1.147 (HxA) [5]; –1.550 (DPA) [7]; –1.997 (TEA) [5]. This has been explained in terms of strong structural effects in systems with TEA, which lead to excess molar internal energies at constant volume (–846 J·mol$^{-1}$ for the propan-1-ol + TEA mixture [5]) which largely differ from the corresponding $H_m^E$ results (–1413 J·mol$^{-1}$ [12]).

We have investigated in detail alkan-1-ol + amine systems by means of different models [1,5-8,13-16]: DISQUAC [17,18], ERAS [1,3], the concentration-concentration structure factor, $S_{CC}(0)$ [19,20], or the Kirkwood-Buff formalism [21]. In addition, we have reported data on $V_m^E$ [5,7,8], vapor-liquid equilibria [22] or viscosity [23-25]. More recently, we have determined permittivities, $\varepsilon_r$, and refractive indices, $n_D$, for alkan-1-ol + cyclohexylamine [26], or + HxA [27], or + DPA [28] mixtures. As a continuation, and in order to investigate the influence of the shape of TEA molecules on dielectric properties, here we provide $\varepsilon_r$ and $n_D$ measurements for methanol, or propan-1-ol, or butan-1-ol, or pentan-1-ol, or heptan-1-ol + TEA systems at (293.15-303.15) K. In addition, and as in previous applications [26-28], the $\varepsilon_r$ and $n_D$ results are used to investigate the systems using the Kirkwood-Fröhlich theory [29-32].

This type of studies is relevant for a better understanding of non-covalent interactions, i.e. hydrogen bonding. Hydrogen bonding leads to cooperative effects which are crucial in supramolecular chemistry and biochemistry [33,34]. Thus, such effects are essential for the characterization of association of molecules in the condensed phase [35,36] or of the DNA molecule [37]. Mixtures with amines are of particular interest, since the disruption of amino acids releases amines and proteins that are usually bound to DNA polymers contain several amine groups [38]. They are also used for $CO_2$ capture [39]. Finally, we remark that many of the ions of the technically important ionic liquids include amine groups [40].

## 2. Experimental

### *2.1 Materials*

Pure compounds were used in the experiments without further purification. Information about their source and purity is shown in Table 1. Water content of the employed chemicals was determined using the Karl-Fischer method. The relative standard uncertainty of these measurements is estimated to be 0.025. In the present work, the water content was ignored along calculations. Values of $\varepsilon_r$ at 1 MHz, density ($\rho$) and $n_D$ at the working temperatures and the dipole moments ($\mu$) of the pure compounds are collected in Table 2. Comparison with literature values reveals a good agreement.

### *2.2 Apparatus and procedure*

Binary mixtures were prepared by mass in small vessels of about 10 $cm^3$ using an analytical balance Sartorius MSU125p (weighing accuracy $10^{-8}$ kg), correcting the weighings for buoyancy effects. The standard uncertainty in the mole fraction is 0.0010. Molar quantities were calculated using the relative atomic mass Table of 2015 issued by the Commission on Isotopic Abundances and Atomic Weights (IUPAC) [41]. Pure liquids were stored with 4 Å molecular sieves (except methanol, because measurements were affected) in order to minimize the effects of the interaction with air components. The measurement cell (see below) was completely filled with the samples and appropriately closed to avoid their evaporation. The density of the pure compounds was measured along the experiments, remaining constant within the experimental uncertainty.

Temperatures were measured with Pt-100 resistances calibrated according to the ITS-90 scale of temperature, using the triple point of water and the melting point of Ga as reference points. The standard uncertainty of this quantity is 0.01 K for $\rho$ measurements, and 0.02 K for $\varepsilon_{\mathrm{r}}$ and $n_{\mathrm{D}}$ measurements.

A RFM970 refractometer from Bellingham + Stanley was used for the experimental $n_{\mathrm{D}}$ determination. This device exploits the optical detection of the critical angle at the wavelength of the sodium D line (589.3 nm). A temperature stability of 0.02 K is guaranteed by Peltier modules. The calibration of the refractometer was performed using 2,2,4-trimethylpentane and toluene at (293.15 – 303.15) K, following the recommendations by Marsh [42]. The standard uncertainty of $n_{\mathrm{D}}$ is 0.00008.

Densities were obtained using a vibrating-tube densimeter and sound analyzer Anton Paar DSA 5000, which is automatically thermostated within 0.01 K. The calibration procedure has been described elsewhere [43]. The relative standard uncertainty of the $\rho$ measurements is 0.0012.

The experimental device to determine $\varepsilon_{\mathrm{r}}$ consists of a 16452A cell –parallel-plate capacitor made of Nickel-plated cobalt (54% Fe, 17% Co, 29% Ni) with a ceramic insulator (alumina, $Al_2O_3$)–, which is filled with a sample volume of $\approx 4.8$ cm$^3$ and connected by a 16048G test lead to a precision impedance analyzer 4294A, all of them from Agilent. The cell is immersed in a thermostatic bath LAUDA RE304, with a temperature stability of 0.02 K. Details about the equipment configuration and calibration are given elsewhere [44]. The relative standard uncertainty of the $\varepsilon_{\mathrm{r}}$ measurements (i.e. the repeatability) is 0.0001. The total relative standard uncertainty of $\varepsilon_{\mathrm{r}}$ was estimated to be 0.003 from the differences between our data and values available in the literature, in the range of temperature (288.15 – 333.15) K, for the following pure liquids: water, benzene, cyclohexane, hexane, nonane, decane, dimethyl carbonate, diethyl carbonate, methanol, propan-1-ol, pentan-1-ol, hexan-1-ol, heptan-1-ol, octan-1-ol, nonan-1-ol and decan-1-ol.

## 3. Results

The volume fraction of component *i*, $\phi_i$, is calculated as $\phi_i = x_i V_{\mathrm{m}i}^* / \left( x_1 V_{\mathrm{m}1}^* + x_2 V_{\mathrm{m}2}^* \right)$, where $x_i$ is the mole fraction of component $i$ and $V_{\mathrm{m}i}^*$ is its molar volume. The derivative $\left( \partial \varepsilon_{\mathrm{r}} / \partial T \right)_p$ was calculated at 298.15 K as the slope of a linear regression of experimental $\varepsilon_{\mathrm{r}}$ values in the range (293.15 – 303.15) K. For an ideal mixture at the same temperature and pressure as the mixture under study, the relative permittivity, $\varepsilon_{\mathrm{r}}^{\mathrm{id}}$, the derivative $\left[\left(\partial\varepsilon_{\mathrm{r}}/\partial T\right)_p\right]^{\mathrm{id}}$, and the refractive index, $n_{\mathrm{D}}^{\mathrm{id}}$, are given by [45,46]:

$$\varepsilon_{\mathrm{r}}^{\mathrm{id}} = \phi_1\varepsilon_{\mathrm{r}1}^{*} + \phi_2\varepsilon_{\mathrm{r}2}^{*} \qquad (1)$$

$$n_{\mathrm{D}}^{\mathrm{id}} = \left[\phi_1\left(n_{\mathrm{D}1}^{*}\right)^2 + \phi_2\left(n_{\mathrm{D}2}^{*}\right)^2\right]^{1/2} \qquad (2)$$

$$\left[\left(\frac{\partial\varepsilon_{\mathrm{r}}}{\partial T}\right)_p\right]^{\mathrm{id}} = \left(\frac{\partial\varepsilon_{\mathrm{r}}^{\mathrm{id}}}{\partial T}\right)_p \qquad (3)$$

where $\varepsilon_{\mathrm{r}i}^{*}$ and $n_{\mathrm{D}i}^{*}$ denote the relative permittivity and the refractive index of pure species $i$, and $\left(\partial\varepsilon_{\mathrm{r}}^{\mathrm{id}}/\partial T\right)_p$ is calculated from linear regressions as already mentioned. The corresponding excess functions, $F^{\mathrm{E}}$, are calculated from these according to the equation:

$$F^{\mathrm{E}} = F - F^{\mathrm{id}} \quad , \quad F = \varepsilon_{\mathrm{r}},\ n_{\mathrm{D}},\ \left(\frac{\partial\varepsilon_{\mathrm{r}}}{\partial T}\right)_p \qquad (4)$$

Table 3 collects $\phi_1$, $\varepsilon_{\mathrm{r}}$ and $\varepsilon_{\mathrm{r}}^{\mathrm{E}}$ values of alkan-1-ol (1) + TEA (2) systems as functions of $x_1$, in the temperature range (293.15 – 303.15) K. Table 4 contains the experimental $x_1$, $\phi_1$, $n_{\mathrm{D}}$ and $n_{\mathrm{D}}^{\mathrm{E}}$ values at the same conditions. The data of $\left[\left(\partial\varepsilon_{\mathrm{r}}/\partial T\right)_p\right]^{\mathrm{E}} = \left(\partial\varepsilon_{\mathrm{r}}^{\mathrm{E}}/\partial T\right)_p$ at 298.15 K are collected in Table S1 (supplementary material).

The $F^{\mathrm{E}}$ data were fitted to a Redlich-Kister equation [47] by unweighted linear least-squares regressions:

$$F^{\mathrm{E}} = x_1\left(1-x_1\right)\sum_{i=0}^{k-1} A_i\left(2x_1-1\right)^i \qquad (5)$$

The number, $k$, of appropriate coefficients for each system, property and temperature has been determined by the application of an F-test of additional term [48] at a 99.5% confidence level. Table 5 includes the parameters $A_i$ obtained, and the standard deviations, $\sigma\left(F^{\mathrm{E}}\right)$, defined by:

$$\sigma\left(F^{\mathrm{E}}\right) = \left[\frac{1}{N-k}\sum_{j=1}^{N}\left(F_{\mathrm{cal},j}^{\mathrm{E}} - F_{\mathrm{exp},j}^{\mathrm{E}}\right)^2\right]^{1/2} \qquad (6)$$

where the index $j$ takes values for each of the $N$ experimental data $F_{\mathrm{exp},j}^{\mathrm{E}}$, and $F_{\mathrm{cal},j}^{\mathrm{E}}$ is the corresponding value of the excess property $F^{\mathrm{E}}$ calculated from equation (5).

Values of $\varepsilon_r^E$, $\left(\partial\varepsilon_r^E / \partial T\right)_p$ and $n_D^E$ versus $\phi_1$ of alkan-1-ol + TEA systems at 298.15 K are plotted in Figures 1, 2 and 3 respectively with their corresponding Redlich-Kister regressions. Data on $n_D$ are plotted in Figure S1 (supplementary material).

# 4. Discussion

The present discussion is concerned with binary mixtures including 1-alkanol (component 1, $i$ = 1) and an organic solvent (component 2, usually amine, $i$ = 2). Unless stated otherwise, the below values of the dielectric properties and their corresponding excess functions are referred to $T$ = 298.15 K and $\phi_1 = 0.5$. On the other hand, $n$ will stand for the number of C atoms of the alkan-1-ol.

## 4.1. Relative permittivities

The magnitude of $\varepsilon_r$ for a liquid system is determined by a number of factors, such as the permanent dipole moments and polarizabilities of its molecules, the nature of the liquid structure and collective dynamics. Figure 4 shows our $\varepsilon_r(\phi_1)$ results for systems with methanol or heptan-1-ol and an isomeric amine TEA, HxA [27], or DPA [28]. We note that, at any composition, $\varepsilon_r(\phi_1)$ values for the mixtures with TEA are lower, which indicates that a weakening of the dielectric polarization of the system is produced with regards to that of solutions with linear isomeric amines. This may be ascribed to the dipole moment of TEA ($2.202\cdot10^{-30}$ C·m [49]), which is lower than the dipole moments of HxA ($4.336\cdot10^{-30}$ C·m [50]) or DPA ($3.669\cdot10^{-30}$ C·m [51]). Interestingly, there is a range of concentrations, which depends on the system components, where small negative differences $\varepsilon_r(\phi_1)$(HxA) − $\varepsilon_r(\phi_1)$(DPA) (for a fixed alkan-1-ol) are encountered (Figure 4). Therefore, in those regions, the effective dipole moments of the multimers formed by unlike molecules upon mixing are lower in the case of HxA-containing systems, probably due to the existence of cyclic species. Outside of the mentioned range of compositions, $\varepsilon_r(\phi_1)$(DPA) < $\varepsilon_r(\phi_1)$(HxA), in agreement with the lower dipole moment of DPA. This can be better visualized in Figure 5, where we have eliminated volume effects present in the permittivity by representing the molar susceptibility, $\chi_m = (\varepsilon_r - 1)V_m$ ($V_m$, molar volume of the mixture), of these mixtures vs $\phi_1$. The quantity $\chi_m$ is useful to compare the response of different liquids given a value of the equilibrium electric field, because it is proportional to the macroscopic dipole moment resulting from a fixed amount (1 mol) of molecules. The $\phi_1$ dependence of $\chi_m$ and $\varepsilon_r$ are very different for the methanol + DPA or + HxA systems (Figures 4 and 5). In fact, there is a more or less large concentration range where $\chi_m$ slowly increases, i.e., where the molar macroscopic dipole moment remains nearly unchanged.

### 4.2. Excess relative permittivities

It is known that the rupture of interactions between molecules of the same species upon mixing provides a negative contribution to $\varepsilon_r^E$. For example, $\varepsilon_r^E$ (heptane) = –1.075 ($n$ = 3), –2.225 ($n$ = 4), –2.525 ($n$ = 5), –2.875 ($n$ = 7), –1.775 ($n$ = 10) [24,52,53] (Figure 6). For the system methanol + heptane, a partial immiscibility region appears [54]. These rather large and negative values can be ascribed to the disruption of the alcohol network along the mixing process. The creation of new interactions between unlike molecules along this process leads to the formation of multimers whose molecular structure is determinant to provide a more or less effective impact on the macroscopic response to an electric field. If the mentioned multimers are linear chains, the contribution to $\varepsilon_r^E$ is positive. In contrast, if cyclic species are created, the contribution to $\varepsilon_r^E$ is negative. The $\varepsilon_r^E$ values of alkan-1-ol + TEA mixtures are: 0.074 ($n$ = 1), –1.807 ($n$ = 3), –1.980 ($n$ = 4), –1.874 ($n$ = 5), –1.435 ($n$ = 7) (this work), –0.593 ($T$ = 293.15 K) ($n$ = 10); –0.048 ($T$ = 293.15 K) ($n$ = 12); [55] (Figure 6). The comparison of these results for $n \geq 4$ with the lower values given above for alkan-1-ol + heptane systems reveals that the creation of the new (alkan-1-ol)-TEA interactions contributes positively to the $\varepsilon_r^E$ of the mixture. In addition, positive $\varepsilon_r^E$ values are encountered for the methanol-containing system. An important result is that, for systems with $n$ = 3, $\varepsilon_r^E$ (TEA) < $\varepsilon_r^E$ (heptane). This suggests that TEA is an effective breaker of the alkanol self-association, and that the interactions between unlike molecules do not compensate enough the large negative contribution to $\varepsilon_r^E$ from the disruption of (propan-1-ol)-(propan-1-ol) interactions. The variation of $\varepsilon_r^E$ with the chain length of the alkan-1-ol follows the order: methanol > propan-1-ol > butan-1-ol < pentan-1-ol < heptan-1-ol < decan-1-ol < dodecan-1-ol. In other words, $\varepsilon_r^E$ decreases to a minimum and then increases again. Such a trend is similar to those encountered for alkan-1-ol + heptane (see above), + HxA [27], + DPA [28] or + cyclohexylamine [25,26] systems. The observed $\varepsilon_r^E$ dependence on $n$ has been explained in terms of a weaker and lower self-association of longer alkanols. In the case of alkan-1-ol + amine systems, one must also take into account that the solvation between molecules of different species decreases when $n$ is increased [1,5,7,8,56]. Thus, the mixture polarization shows a weaker variation with $n$ when longer alkan-1-ols are involved, since they are characterized by a lower self-association and the related solvation effects are also less relevant. Consequently, and as in previous studies, $\varepsilon_r^E$ shows a sharper dependence for low $n$ values [27,28]. Other available data on $\varepsilon_r^E$ for the dodecan-1-ol + TEA system at different temperatures [57] should be taken with caution, as they are rather scattered and the corresponding curves are S-shaped, with positive values at higher concentrations of the alkan-1-ol which increase in line with the temperature [57]. This needs further experimental confirmation.

For a given alkan-1-ol, $\varepsilon_r^E$(DPA) [28] > $\varepsilon_r^E$(HxA) [27] > $\varepsilon_r^E$(TEA) (Figure 6). This variation is similar to, although stronger than, the observed change for $\varepsilon_r$. For a better understanding of systems containing TEA, we start examining alkan-1-ol + linear primary or secondary amine systems. A literature survey shows that $\varepsilon_r^E$(propan-1-ol + DPA) = –0.246 [28] > $\varepsilon_r^E$(propan-1-ol + HxA) = –0.96 [27] > $\varepsilon_r^E$(propan-1-ol + propan-1-amine) = –1.99 [58] and that $\varepsilon_r^E$(butan-1-ol + DPA) = –0.715 [28] > $\varepsilon_r^E$ (butan-1-ol + HxA) = –1.424 [27] > $\varepsilon_r^E$(butan-1-ol + butan-1-amine) = –2.87 [58]. Since solvation effects are expected to be more relevant in systems involving amines where the amine group is less sterically hindered (propan-1-amine, butan-1-amine), one can conclude that mixtures characterized by larger solvation effects show more negative $\varepsilon_r^E$ values. The same trend is observed when comparing, at 303.15 K, $\varepsilon_r^E$ results for propan-1-ol + primary aromatic amine, aniline, (–2.07) [59] or + secondary aromatic amine, *N*-methylaniline, (–1.27) [60]. This behavior can be explained taking into account that larger solvation effects imply a decreased number of interactions between like molecules and, therefore, a more negative contribution to $\varepsilon_r^E$ from the disruption of interactions between like molecules, particularly between alkanol molecules. In the case of amine mixtures, cyclic species may be more probable in mixtures containing amines with the characteristic group less sterically hindered. We must now remark that systems with TEA deviate from this picture. This can be ascribed to the globular shape of TEA molecules, which makes them better breakers of the alkan-1-ol self-association (see above). In fact, the volume fraction at which minimum $\varepsilon_r^E$ values are measured changes in the sequence DPA < HxA < TEA for mixtures with shorter alkan-1-ols. Thus, $\phi_1$($n$ = 4) = 0.3183 (DPA; $\varepsilon_r^E$ = –0.896) [28] < 0.4138 (HxA; $\varepsilon_r^E$ = –1.428) [27] < 0.4969 (TEA; $\varepsilon_r^E$ = –1.964). For $n$ = 7, the alcohol self-association becomes less relevant, and the minimum $\varepsilon_r^E$ values are encountered at similar volume fractions for HxA or TEA mixtures, although these concentrations are still higher than for the DPA solution, e.g. $\phi_1$($n$ = 7) = 0.5003 (DPA; $\varepsilon_r^E$ = –0.793) [28] < 0.5982 (TEA; $\varepsilon_r^E$ = –1.455). The fact that the $\varepsilon_r^E$ curves of HxA systems are skewed to higher $\phi_1$ values than those of mixtures with DPA supports our previous statement about that higher solvation effects lead to a more important breaking of the alcohol network upon mixing. We complete the present analysis as follows. (i) According to the ERAS model, the equilibrium constants, $K_{AB}$, change in the order HxA > DPA > TEA in systems with a given alkan-1-ol [5-7]. For example, $K_{AB}$ (methanol) = 2500 (HxA) > 2450 (DPA) > 620 (TEA). That is, solvation effects are less important in mixtures with TEA, in agreement with the fact that the amine group becomes more sterically hindered in the same sequence [56]. (ii) The $S_{CC}(0)$ function is a quantity which allows to study the fluctuations in the number of molecules of a binary mixture regardless of the components, the fluctuations in the mole fraction and the cross fluctuations. It is defined by [19,20]:

$$S_{CC}(0) = \frac{x_1x_2}{1+\frac{x_1x_2}{RT}\left(\frac{\partial^2 G_m^E}{\partial x_1^2}\right)_{T,p}} = \frac{x_1x_2}{D} \quad (7)$$

where $D = 1 + (x_1x_2/RT)\left(\partial^2 G_m^E/\partial x_1^2\right)_{T,p}$. For ideal mixtures, $G_m^{E,id}$ = 0 (excess Gibbs energy of the ideal mixture); $D^{id}$ = 1 and $S_{CC}(0)$ = $x_1x_2$. From stability conditions, $S_{CC}(0)$ > 0. If a system is close to phase separation, $S_{CC}(0)$ must be large and positive ($\infty$, if the mixture presents a miscibility gap). In the case of compound formation between components, $S_{CC}(0)$ must be very low (0, in the limit). Therefore, $S_{CC}(0) > x_1x_2$ ($D < 1$) indicates that the dominant trend in the system is homocoordination (separation of the components), and the mixture is then less stable than the ideal. If $0 < S_{CC}(0) < x_1x_2 = S_{CC}(0)^{id}$, ($D > 1$), the fluctuations in the system have been removed, and the dominant trend in the solution is heterocoordination (compound formation). In such a case, the system is more stable than ideal. We have shortly applied this formalism to methanol + HxA, or + DPA, or + TEA systems at 298.15 K calculating $G_m^E$ by means of the DISQUAC model with interaction parameters for the OH/amine contacts previously determined [1,6]. At equimolar composition, we have obtained: $S_{CC}(0)$ = 0.165 (HxA) < 0.201 (DPA) < 0.341 (TEA). This means that heterocoordination is dominant in the systems with HxA or DPA, while homocoordination is prevalent in the TEA mixture. It is in full agreement with the variation of the $K_{AB}$ constants given above, and with available $G_m^E$ data for methanol + amine mixtures. Thus, $G_m^E$ (methanol)/J·mol$^{-1}$ = –799 (butan-1-amine, 348.15 K) [2], 284 (TEA, 303.15 K) [9]. (iii) Interestingly, viscosity data show that values of $\Delta\eta$ (= $\eta - x_1\eta_1 + x_2\eta_2$; where $\eta$ is the mixture viscosity and $\eta_i$ is the viscosity of component $i$) become more negative in alkan-1-ol mixtures when DPA is replaced by TEA. For example, at 303.15 K and equimolar composition, $\Delta\eta$ (DPA)/10$^{-3}$·Pa·s = –0.142 (propan-1-ol) < –0.259 (butan-1-ol) [61] and $\Delta\eta$ (TEA)/10$^{-3}$·Pa·s = –0.328 (propan-1-ol) [62] < –0.612 (butan-1-ol) [63]. Therefore, the mixture fluidization becomes more relevant in solutions with TEA. This is not only explained by the lower solvation effects present in such systems, but also because a larger number of interactions between alkan-1-ol molecules are broken along the mixing process. Interestingly, at 298.15 K and equimolar composition, $\Delta\eta$ (propan-1-ol)/ $10^{-3}$·Pa·s = –0.305 (propan-1-amine) [64] < –0.253 (butan-1-amine) [65] and $\Delta\eta$ (butan-1-ol)/ $10^{-3}$·Pa·s = –0.460 (propan-1-amine) [64] < –0.280 (butan-1-amine) [65]. It seems that for alkan-1-ol + linear primary amine systems including compounds of similar size and shape, $\Delta\eta$ is lower for the solutions with larger solvation effects. This trend is still valid for mixtures including short chain alkan-1-ols and DPA. For instance, at 303.15 K and $x_1$ = 0.5, the values $\Delta\eta$ (propan-1-amine)/ $10^{-3}$·Pa·s = –0.252 (propan-1-ol); –0.320 (butan-1-ol) [66] are lower than the results given above for alkan-1-ol + DPA systems.

### 4.3. Temperature dependence of the permittivity

Values of $\left(\partial\varepsilon_{\mathrm{r}}^{*}/\partial T\right)_{p}$ of pure compounds used in this work are negative, as it is usual for normal liquids. Pure TEA shows a very low absolute value of this derivative, $\left(\partial\varepsilon_{\mathrm{r}}^{*}/\partial T\right)_{p}$ = -0.004 $K^{-1}$, since TEA is not self-associated and has a low $\varepsilon_{\mathrm{r}}^{*}$ value (= 2.419). Thus, the increase of thermal agitation hardly modifies the liquid structure. On the other hand, values of $\left(\partial\varepsilon_{\mathrm{r}}/\partial T\right)_{p}$ of TEA systems are higher than for pure alkanols (e.g., for pentan-1-ol $\left(\partial\varepsilon_{\mathrm{r}}^{*}/\partial T\right)_{p}$ = –0.117 $K^{-1}$ < $\left(\partial\varepsilon_{\mathrm{r}}/\partial T\right)_{p}$ = –0.034; Figure 7, Table S2). This can be explained as follows. (i) The contribution to $\left(\partial\varepsilon_{\mathrm{r}}/\partial T\right)_{p}$ related to the breaking of TEA-TEA interactions when $T$ is increased is practically negligible (see above); (ii) The enthalpy of hydrogen bonds between alkan-1-ol molecules is larger than that corresponding to alkan-1-ol-TEA interactions. Thus, in the framework of the ERAS model, $\Delta h_{\mathrm{AB}}^{*}$ (TEA)/ kJ·mol$^{-1}$ = –35.3 (methanol); –30.5 (heptan-1-ol) [5], while the enthalpies between alkan-1-ol molecules are –25 kJ·mol$^{-1}$ [4,5,7,67]. Therefore, one can expect that the number of (alkan-1-ol)-TEA interactions broken when the temperature is increased is lower than the number of disrupted alkanol-alkanol interactions. This leads to a lower $\varepsilon_{\mathrm{r}}$ decrease when $T$ is increased in comparison to that produced in pure alkan-1-ols. The variation of $\left(\partial\varepsilon_{\mathrm{r}}/\partial T\right)_{p}$ with $n$ can be explained in similar terms. On the other hand, for mixtures with a given alkan-1-ol, $\left(\partial\varepsilon_{\mathrm{r}}/\partial T\right)_{p}$ changes in the order TEA > HxA > DPA (Figure 7). The $\varepsilon_{\mathrm{r}}$ values vary in the opposite sequence (Figure 4). That is, the structure of mixtures characterized by a higher dielectric polarization is more sensitive to temperature changes. In addition, $\left(\partial\varepsilon_{\mathrm{r}}^{*}/\partial T\right)_{p}$ / $K^{-1}$ = –0.004 (TEA) > –0.0098 (HxA) > –0.012 (DPA).

Finally, we note that $\left(\partial\varepsilon_r^E/\partial T\right)_p$ / $K^{-1}$ can show negative or positive values (Figures 2 and S2): –0.003 ($n$ = 1), 0.017 ($n$ = 3), 0.022 ($n$ = 4), 0.025 ($n$ = 5), 0.027 ($n$ = 7). The negative value is encountered only for the methanol solution, for which the effects from alkan-1-ol self-association and solvation between unlike molecules are more relevant, leading to a network that is more difficult to break with the increasing of temperature when compared with the ideal mixture. DPA and HxA systems behave similarly [27,28] but, since association/solvation effects are stronger, the corresponding values are more negative.

### 4.4. Molar refraction

The molar refraction or molar refractivity is defined by the Lorentz-Lorenz equation [30,32]:

$$R_m = \frac{n_D^2 - 1}{n_D^2 + 2} V_m = \frac{N_A \alpha_e}{3\varepsilon_0} \qquad (8)$$

where $N_A$ is Avogadro's constant and $\varepsilon_0$ denotes the vacuum permittivity. $R_m$ is related to dispersive interactions [32,68], since it is proportional to the average electronic contribution to the polarizability, $\alpha_e$, from one molecule in a macroscopic sphere of liquid [30,32]. The $R_m$ / $10^{-6}\cdot m^3\cdot mol^{-1}$ values for alkan-1-ol + TEA mixtures are (at equimolar composition): 20.7 ($n$ = 1), 25.4 ($n$ = 3), 27.7 ($n$ = 4), 30.0 ($n$ = 5), 34.7 ($n$ = 7). That is, dispersive interactions are more relevant in mixtures with longer alkan-1-ols. Interestingly, the $R_m$ values are very similar to those of alkan-1-ol + HxA or + DPA systems [27,28], and one can conclude that these mixtures differ essentially by solvation effects. The application of the ERAS model [5,7] to alkan-1-ol + HxA, or + DPA, or + TEA systems shows that these mixtures are characterized by the same small physical parameter and that they differ in the parameters $K_{AB}$, $\Delta h_{AB}^*$ and $\Delta v_{AB}^*$.

### 4.5. Aromaticity effect

The available $\varepsilon_r^E$ data in the literature for alkan-1-ol + pyridine systems [69,70] indicate that they are higher than those of the TEA solutions. Thus, $\varepsilon_r^E$ (methanol) = 2.85 (pyridine) [69] > 0.074 (TEA, this work); and, at 303.15 K, $\varepsilon_r^E$ (propan-1-ol) = 0.10 [70] (pyridine) > –1.807 (TEA, this work). Therefore, cooperative effects which lead to an increase of the dielectric polarization of the mixture are more relevant in the systems with pyridine, probably because the amine group is less sterically hindered in the aromatic amine and the creation of multimers, formed by unlike molecules, with larger effective dipole moments is favored. This effect predominates over the larger negative contribution to $\varepsilon_r^E$ from the breaking of dipolar interactions between pyridine molecules. Note that the dipole moment of pyridine (2.37 D, [71]) is much higher than the dipole moment of TEA. It is remarkable that the behavior of alkan-1-ol

+ HxA, or + aniline systems is the opposite, and $\varepsilon_r^E$ values are more negative for the solutions involving aniline [27]. That is, the large negative contribution to $\varepsilon_r^E$ from the disruption of aniline-aniline interactions predominates. The mentioned interactions are much stronger than those between pyridine molecules, as it is shown by the upper critical solution temperatures (UCST) of their mixtures with *n*-alkanes. For example. UCST/K= 268.7 (pyridine + dodecane) [72] < 343.1 (aniline + heptane) [73].

### 4.6. Kirkwood-Fröhlich model

In the Kirkwood-Fröhlich model, the fluctuations of the dipole moment in the absence of electric field are treated as the basis to obtain relations involving the relative permittivity. It is a local-field model in which the molecules are assumed to be in a spherical cavity of an infinitely large piece of dielectric and the induced contribution to the polarizability is treated macroscopically through its relation to $\varepsilon_r^\infty$ (the value of the permittivity at a high frequency at which only the induced polarizability contributes). The local field takes into account long-range dipolar interactions by considering the outside of the cavity as a continuous medium of permittivity $\varepsilon_r$ . Short-range interactions are introduced by the so-called Kirkwood correlation factor, $g_K$, which provides information about the deviations from randomness of the orientation of a dipole with respect to its neighbors. This is an important parameter, as it provides information about specific interactions in the liquid state. For a mixture of polar liquids, $g_K$ can be determined, in a one-fluid model approach [29], from macroscopic physical properties according to the expression [29-32]:

$$g_K = \frac{9k_B T V_m \varepsilon_0 (\varepsilon_r - \varepsilon_r^\infty)(2\varepsilon_r + \varepsilon_r^\infty)}{N_A \mu^2 \varepsilon_r (\varepsilon_r^\infty + 2)^2} \quad (9)$$

Here, $k_B$ is Boltzmann's constant; $N_A$, Avogadro's constant; $\varepsilon_0$, the vacuum permittivity; and $V_m$, the molar volume of the liquid at the working temperature, *T*. For polar compounds, $\varepsilon_r^\infty$ is estimated from the relation $\varepsilon_r^\infty = 1.1 n_D^2$ [74]. $\mu$ represents the dipole moment of the solution, estimated from the equation [29]:

$$\mu^2 = x_1 \mu_1^2 + x_2 \mu_2^2 \quad (10)$$

where $\mu_i$ stands for the dipole moment of component *i* (= 1,2). Calculations have been performed using smoothed values of $V_m^E$ [5], $n_D^E$ (this work) and $\varepsilon_r^E$ (this work) at $\Delta x_1$ = 0.01. The source and values of $\mu_i$ are collected in Table 2.

We compare the $g_K$ curves obtained from methanol or heptan-1-ol + isomeric amine in Figure 8. Except for values of $\phi_1$ very close to zero, where the structure of the mixture is basically that of the pure amine, it is found that $g_K$(DPA) > $g_K$(TEA) > $g_K$(HxA). In order to examine these results, we provide some $g_K$ values for alkan-1-ol + amine mixtures. Thus, $g_K$ (propan-1-ol) = 2.72 (DPA) > 2.32 (HxA) > 1.87 (propan-1-amine), and $g_K$(butan-1-ol) = 2.60 (DPA) > 2.16 (HxA) > 1.72 (butan-1-amine). In addition, $g_K$(propan-1-ol, 303.15 K) = 1.54 (aniline) < 1.71 (*N*-methylaniline). This points out that parallel alignment of molecular dipoles has a lower weight in those systems characterized by larger solvation effects and, according to our previous description of $\varepsilon_r^E$, these cooperative effects will lead to lower polarization of the mixture. This underlines the lower contribution to the mixture structure from alkanol-alkanol interactions in systems with larger solvation effects, and suggests the presence of cyclic species in such systems. The $g_K$ results for the methanol + DPA mixture deserve a comment. We note that $g_K$ rapidly increases with $\phi_1$, and that it is nearly constant from $\phi_1$ = 0.5 and very close to the value of the neat alcohol. This might occur because the contribution to the mixture polarization arising from interactions between alcohol molecules also increases rapidly with $\phi_1$ in such a way that interactions between unlike molecules contribute to $g_K$ to a lower extent. It is remarkable that $g_K$ changes more smoothly with $\phi_1$ for the methanol + HxA system, in agreement with our analysis of $\varepsilon_r^E$ results. For alkan-1-ol + TEA systems, $g_K$ = 2.73 ($n$ = 1), 2.47 ($n$ = 3), 2.38 ($n$ = 4), 2.30 ($n$ = 5), 2.13 ($n$ = 7) (see Figure S3). For $n$ = 1, the $g_K$ curve remains nearly constant from $\phi_1 > 0.7$. It is quite clear that TEA mixtures show an intermediate behavior, which could be due to the existence of a higher proportion of shorter linear-like multimers of alkan-1-ol molecules which are less present in the systems with HxA.

The excess Kirkwood correlation factors ($g_K^E = g_K - g_K^{id}$, where $g_K^{id}$ is obtained replacing real by ideal quantities in equation (9)) of alkan-1-ol + TEA mixtures are: –0.03 ($n$ = 1), –0.58 ($n$ = 3), –0.76 ($n$ = 4), –0.85 ($n$ = 5), –0.84 ($n$ = 7). Their curves are plotted in Figure S4. The variation of the minimum of the curves as $n$ increases is not the same as the one encountered for $\varepsilon_r^E$, and it occurs at lower values of $\phi_1$. The model consequently indicates that the variation of the structure of the dipoles in the mixing process is only one of the factors that determine the $\varepsilon_r^E$ minima. The $g_K^E$ values are compared with the corresponding ones of alkan-1-ol + HxA or DPA systems in Figure S5. The trend of $g_K^E$(TEA) is slightly deviated from the parallel behavior of $g_K^E$(HxA) and $g_K^E$(DPA). This underlines the stronger structural effects already mentioned in the former mixtures.

## 5. Conclusions

Measurements of $\varepsilon_r$ and $n_D$ have been reported for the alkan-1-ol + TEA mixtures at (293.15-303.15) K. Positive $\varepsilon_r^E$ results are encountered only for the methanol system at high $\phi_1$ values. $\varepsilon_r^E$ changes in the sequence: methanol > propan-1-ol > butan-1-ol < pentan-1-ol < heptan-1-ol. This variation is similar to alkan-1-ol + HxA, or + DPA or + cyclohexylamine. It has been shown that: (i) (alkan-1-ol)-TEA interactions contribute positively to $\varepsilon_r^E$; (ii) TEA is a good breaker of the alkan-1-ol self-association; (iii) structural effects are relevant for $\varepsilon_r$ data. (iv) the aromaticity effect leads to an increase of the mixture polarization, and it is opposite to the effect encountered when considering alkan-1-ol + HxA, or + aniline mixtures. The application of the Kirkwood-Fröhlich model supports these statements.

## Acknowledgements

F. Hevia and A. Cobos are grateful to Ministerio de Educación, Cultura y Deporte for the grants FPU14/04104 and FPU15/05456 respectively.

Table 1

Sample description.

| Chemical name | CAS Number | Source | Purification method | Purity[a] | Water content[b] |
|---|---|---|---|---|---|
| Methanol | 67-56-1 | Sigma-Aldrich | None | 0.9999 | $10^{-5}$ |
| propan-1-ol | 71-23-8 | Sigma-Aldrich | None | 0.9984 | $10^{-3}$ |
| butan-1-ol | 71-36-3 | Sigma-Aldrich | None | 0.9986 | $10^{-3}$ |
| pentan-1-ol | 71-41-0 | Sigma-Aldrich | None | 0.999 | $2 \cdot 10^{-4}$ |
| heptan-1-ol | 111-70-6 | Sigma-Aldrich | None | 0.998 | $5 \cdot 10^{-4}$ |
| *N,N*-diethylethanamine (TEA) | 121-44-8 | Sigma-Aldrich | None | 0.9999 | $10^{-4}$ |

[a] In mole fraction. By gas chromatography. Provided by the supplier.

[b] In mass fraction. By Karl-Fischer titration; the relative standard uncertainty for water content is 0.025.

Table 2

Dipole moment, $\mu$, of the pure liquids, and their relative permittivity at frequency $\nu = 1$ MHz, $\varepsilon_r^*$, refractive index at the sodium D-line, $n_D^*$, and density, $\rho^*$, at temperature $T$ and pressure $p$ = 0.1 MPa. [a]

| Compound | $\mu$ / $10^{-30}$ C·m | $T$/K | $\varepsilon_r^*$ | | $n_D^*$ | | $\rho^*$ / kg·m$^{-3}$ | |
|---|---|---|---|---|---|---|---|---|
| | | | Exp. | Lit. | Exp. | Lit. | Exp. | Lit. |
| methanol | 5.551 [75] | 293.15 | 33.576 | 33.61 [76] | 1.32863 | 1.32859 [77] | 791.63 | 791.6 [78]<br>791.400 [79] |
| | | 298.15 | 32.624 | 32.62 [76] | 1.32649 | 1.32652 [80] | 786.95 | 786.9 [81]<br>786.884 [82] |
| | | 303.15 | 31.684 | 31.66 [76] | 1.32435 | 1.32457 [83]<br>1.32410 [84] | 782.22 | 782.158 [82] |
| propan-1-ol | 5.434 [75] | 293.15 | 21.150 | 21.15 [85] | 1.38511 | 1.38512 [86] | 803.66 | 803.61 [87] |
| | | 298.15 | 20.469 | 20.42 [85] | 1.38306 | 1.38307 [84] | 799.68 | 799.60 [87] |
| | | 303.15 | 19.799 | 19.75 [85] | 1.38099 | 1.38104 [84] | 795.66 | 795.61 [87] |
| butan-1-ol | 5.384 [75] | 293.15 | 18.201 | 18.19 [85] | 1.39929 | 1.3993 [88] | 809.85 | 809.82 [89]<br>809.8 [90] |
| | | 298.15 | 17.566 | 17.53 [85] | 1.39730 | 1.397336 [91] | 806.06 | 806.06 [89] |
| | | 303.15 | 16.942 | 16.89 [85] | 1.39529 | 1.3953 [92] | 802.22 | 802.2 [90] |
| pentan-1-ol | 5.330 [75] | 293.15 | 15.689 | 15.63 [76] | 1.40993 | 1.40986 [84] | 814.66 | 814.68 [93] |
| | | 298.15 | 15.110 | 15.08 [94] | 1.40794 | 1.40789 [84] | 811.03 | 811.03 [93] |
| | | 303.15 | 14.537 | 14.44 [76] | 1.40592 | 1.40592 [95] | 807.35 | 817.37 [93] |
| heptan-1-ol | 5.280 [75] | 293.15 | 12.005 | 11.54 [96] | 1.42433 | 1.42433 [97] | 822.37 | 822.3 [98] |
| | | 298.15 | 11.504 | 11.45 [94] | 1.42236 | 1.42240 [97] | 818.90 | 818.81 [99] |
| | | 303.15 | 11.013 | 11.07 [100] | 1.42041 | 1.42047 [95]<br>1.42048 [97] | 815.37 | 815.3 [98] |
| TEA | 2.202 [49] | 293.15 | 2.440 | 2.43 [101]<br>2.450 [102]<br>2.46 [103] | 1.40044 | 1.40040 [104]<br>1.1004 [101]<br>1.400333 [105] | 727.38 | 726.6 [101]<br>727.6 [71] |
| | | 298.15 | 2.419 | 2.42 [103]<br>2.404 [106] | 1.39775 | 1.39825 [104]<br>1.3983 [107] | 722.76 | 723.06 [71] |
| | | 303.15 | 2.398 | 2.387 [106]<br>2.41 [103] | 1.39503 | 1.39555 [104]<br>1.3955 [107] | 718.11 | 717.9 [108] |

[a] The standard uncertainties are: $u(T)$ = 0.02 K (for $\rho^*$ measurements, $u(T)$ = 0.02 K); $u(p)$ = 1 kPa; $u(\nu)$ = 20 Hz; $u(n_D^*)$ = 0.00008. The relative standard uncertainties are: $u_r(\rho^*)$ = 0.0012, $u_r(\varepsilon_r^*)$ = 0.003.

Table 3

Volume fractions of alkan-1-ol, $\phi_1$, relative permittivities at frequency $\nu$ = 1 MHz, $\varepsilon_r$, and excess relative permittivities at $\nu$ = 1 MHz, $\varepsilon_r^E$, of alkan-1-ol (1) + TEA (2) liquid mixtures as functions of the mole fraction of the alkan-1-ol, $x_1$, at temperature $T$ and pressure $p$ = 0.1 MPa. [a]

| $x_1$ | $\phi_1$ | $\varepsilon_r$ | $\varepsilon_r^E$ | $x_1$ | $\phi_1$ | $\varepsilon_r$ | $\varepsilon_r^E$ |
|---|---|---|---|---|---|---|---|
| | | methanol (1) + TEA (2) ; $T$/K = 293.15 | | | | | |
| 0.0000 | 0.0000 | 2.442 | | 0.5979 | 0.3020 | 11.168 | –0.676 |
| 0.0580 | 0.0176 | 2.801 | –0.189 | 0.7096 | 0.4155 | 15.117 | –0.261 |
| 0.0878 | 0.0272 | 3.038 | –0.251 | 0.7992 | 0.5366 | 19.374 | 0.225 |
| 0.1593 | 0.0523 | 3.575 | –0.495 | 0.8445 | 0.6124 | 21.941 | 0.433 |
| 0.1981 | 0.0671 | 3.939 | –0.592 | 0.8997 | 0.7230 | 25.536 | 0.584 |
| 0.2880 | 0.1053 | 4.938 | –0.782 | 0.9515 | 0.8509 | 29.454 | 0.520 |
| 0.3971 | 0.1608 | 6.538 | –0.910 | 0.9856 | 0.9522 | 32.327 | 0.239 |
| 0.4959 | 0.2225 | 8.475 | –0.894 | 1.0000 | 1.0000 | 33.576 | |
| | | methanol (1) + TEA (2) ; $T$/K = 298.15 | | | | | |
| 0.0000 | 0.0000 | 2.422 | | 0.5979 | 0.3019 | 10.874 | –0.666 |
| 0.0580 | 0.0176 | 2.769 | –0.185 | 0.7096 | 0.4154 | 14.708 | –0.260 |
| 0.0878 | 0.0272 | 2.998 | –0.245 | 0.7992 | 0.5365 | 18.829 | 0.204 |
| 0.1593 | 0.0522 | 3.518 | –0.481 | 0.8445 | 0.6123 | 21.315 | 0.400 |
| 0.1981 | 0.0670 | 3.868 | –0.578 | 0.8997 | 0.7229 | 24.811 | 0.556 |
| 0.2880 | 0.1053 | 4.834 | –0.768 | 0.9515 | 0.8509 | 28.615 | 0.494 |
| 0.3971 | 0.1608 | 6.387 | –0.891 | 0.9856 | 0.9522 | 31.414 | 0.234 |
| 0.4959 | 0.2224 | 8.263 | –0.876 | 1.0000 | 1.0000 | 32.624 | |
| | | methanol (1) + TEA (2) ; $T$/K = 303.15 | | | | | |
| 0.0000 | 0.0000 | 2.402 | | 0.5979 | 0.3018 | 10.582 | –0.657 |
| 0.0580 | 0.0176 | 2.736 | –0.181 | 0.7096 | 0.4153 | 14.302 | –0.261 |
| 0.0878 | 0.0272 | 2.958 | –0.240 | 0.7992 | 0.5364 | 18.293 | 0.184 |
| 0.1593 | 0.0522 | 3.461 | –0.470 | 0.8445 | 0.6122 | 20.705 | 0.377 |
| 0.1981 | 0.0670 | 3.799 | –0.565 | 0.8997 | 0.7228 | 24.096 | 0.529 |
| 0.2880 | 0.1052 | 4.732 | –0.750 | 0.9515 | 0.8508 | 27.792 | 0.477 |
| 0.3971 | 0.1607 | 6.236 | –0.872 | 0.9856 | 0.9521 | 30.513 | 0.232 |
| 0.4959 | 0.2224 | 8.054 | –0.860 | 1.0000 | 1.0000 | 31.684 | |
| | | propan-1-ol (1) + TEA (2) ; $T$/K = 293.15 | | | | | |
| 0.0000 | 0.0000 | 2.440 | | 0.6027 | 0.4492 | 8.889 | –1.956 |
| 0.0476 | 0.0262 | 2.686 | –0.244 | 0.6997 | 0.5560 | 11.070 | –1.773 |
| 0.0932 | 0.0524 | 2.954 | –0.466 | 0.7966 | 0.6780 | 13.743 | –1.382 |
| 0.1394 | 0.0801 | 3.248 | –0.691 | 0.8455 | 0.7463 | 15.298 | –1.105 |
| 0.2032 | 0.1206 | 3.718 | –0.978 | 0.9014 | 0.8309 | 17.310 | –0.676 |
| 0.2896 | 0.1797 | 4.459 | –1.343 | 0.9484 | 0.9081 | 19.076 | –0.355 |
| 0.4084 | 0.2706 | 5.770 | –1.733 | 1.0000 | 1.0000 | 21.150 | |
| 0.5043 | 0.3535 | 7.131 | –1.923 | | | | |
| | | propan-1-ol (1) + TEA (2) ; $T$/K = 298.15 | | | | | |
| 0.0000 | 0.0000 | 2.419 | | 0.6027 | 0.4488 | 8.651 | –1.869 |
| 0.0476 | 0.0261 | 2.660 | –0.230 | 0.6997 | 0.5557 | 10.757 | –1.692 |
| 0.0932 | 0.0523 | 2.916 | –0.447 | 0.7966 | 0.6776 | 13.343 | –1.307 |
| 0.1394 | 0.0800 | 3.200 | –0.663 | 0.8455 | 0.7460 | 14.842 | –1.042 |
| 0.2032 | 0.1204 | 3.654 | –0.938 | 0.9014 | 0.8307 | 16.775 | –0.638 |
| 0.2896 | 0.1795 | 4.368 | –1.291 | 0.9484 | 0.9080 | 18.481 | –0.327 |
| 0.4084 | 0.2704 | 5.640 | –1.660 | 1.0000 | 1.0000 | 20.469 | |
| 0.5043 | 0.3532 | 6.954 | –1.840 | | | | |
| | | propan-1-ol (1) + TEA (2) ; *T*/K = 303.15 | | | | | |
| 0.0000 | 0.0000 | 2.398 | | 0.6027 | 0.4485 | 8.417 | –1.785 |
| 0.0476 | 0.0261 | 2.631 | –0.221 | 0.6997 | 0.5553 | 10.450 | –1.611 |
| 0.0932 | 0.0522 | 2.878 | –0.428 | 0.7966 | 0.6773 | 12.945 | –1.239 |
| 0.1394 | 0.0799 | 3.153 | –0.635 | 0.8455 | 0.7458 | 14.392 | –0.984 |
| 0.2032 | 0.1203 | 3.592 | –0.899 | 0.9014 | 0.8305 | 16.243 | –0.607 |
| 0.2896 | 0.1793 | 4.281 | –1.237 | 0.9484 | 0.9078 | 17.887 | –0.308 |
| 0.4084 | 0.2701 | 5.510 | –1.588 | 1.0000 | 1.0000 | 19.799 | |
| 0.5043 | 0.3529 | 6.781 | –1.758 | | | | |
| | | butan-1-ol (1) + TEA (2) ; *T*/K = 293.15 | | | | | |
| 0.0000 | 0.0000 | 2.444 | | 0.6006 | 0.4973 | 8.204 | –2.076 |
| 0.0536 | 0.0359 | 2.727 | –0.283 | 0.6985 | 0.6038 | 10.001 | –1.957 |
| 0.1102 | 0.0753 | 3.052 | –0.579 | 0.8023 | 0.7275 | 12.371 | –1.536 |
| 0.1536 | 0.1067 | 3.335 | –0.790 | 0.8438 | 0.7804 | 13.458 | –1.283 |
| 0.2025 | 0.1431 | 3.670 | –1.029 | 0.8910 | 0.8432 | 14.801 | –0.929 |
| 0.2941 | 0.2151 | 4.453 | –1.380 | 0.9475 | 0.9223 | 16.509 | –0.468 |
| 0.3957 | 0.3011 | 5.409 | –1.779 | 1.0000 | 1.0000 | 18.201 | |
| 0.5082 | 0.4047 | 6.786 | –2.035 | | | | |
| | | butan-1-ol (1) + TEA (2) ; *T*/K = 298.15 | | | | | |
| 0.0000 | 0.0000 | 2.424 | | 0.6006 | 0.4969 | 7.984 | –1.964 |
| 0.0536 | 0.0359 | 2.696 | –0.272 | 0.6985 | 0.6034 | 9.718 | –1.843 |
| 0.1102 | 0.0752 | 3.010 | –0.553 | 0.8023 | 0.7272 | 11.994 | –1.441 |
| 0.1536 | 0.1065 | 3.285 | –0.752 | 0.8438 | 0.7801 | 13.032 | –1.204 |
| 0.2025 | 0.1429 | 3.608 | –0.980 | 0.8910 | 0.8430 | 14.322 | –0.867 |
| 0.2941 | 0.2148 | 4.365 | –1.312 | 0.9475 | 0.9222 | 15.961 | –0.427 |
| 0.3957 | 0.3007 | 5.288 | –1.689 | 1.0000 | 1.0000 | 17.566 | |
| 0.5082 | 0.4043 | 6.618 | –1.928 | | | | |
| | | butan-1-ol (1) + TEA (2) ; *T*/K = 303.15 | | | | | |
| 0.0000 | 0.0000 | 2.403 | | 0.6006 | 0.4965 | 7.768 | –1.854 |
| 0.0536 | 0.0358 | 2.666 | –0.257 | 0.6985 | 0.6030 | 9.438 | –1.732 |
| 0.1102 | 0.0751 | 2.970 | –0.525 | 0.8023 | 0.7268 | 11.621 | –1.349 |
| 0.1536 | 0.1063 | 3.235 | –0.713 | 0.8438 | 0.7798 | 12.614 | –1.127 |
| 0.2025 | 0.1427 | 3.547 | –0.931 | 0.8910 | 0.8428 | 13.847 | –0.809 |
| 0.2941 | 0.2146 | 4.278 | –1.245 | 0.9475 | 0.9221 | 15.415 | –0.394 |
| 0.3957 | 0.3004 | 5.169 | –1.602 | 1.0000 | 1.0000 | 16.942 | |
| 0.5082 | 0.4039 | 6.449 | –1.826 | | | | |
| | | pentan-1-ol (1) + TEA (2) ; *T*/K = 293.15 | | | | | |
| 0.0000 | 0.0000 | 2.437 | | 0.5968 | 0.5352 | 7.514 | –2.015 |
| 0.0588 | 0.0463 | 2.743 | –0.308 | 0.7072 | 0.6526 | 9.204 | –1.881 |
| 0.1062 | 0.0846 | 3.009 | –0.549 | 0.7960 | 0.7522 | 10.867 | –1.538 |
| 0.1482 | 0.1192 | 3.260 | –0.757 | 0.8517 | 0.8171 | 12.041 | –1.224 |
| 0.2159 | 0.1764 | 3.708 | –1.067 | 0.8939 | 0.8676 | 13.011 | –0.923 |
| 0.3011 | 0.2510 | 4.361 | –1.402 | 0.9465 | 0.9323 | 14.330 | –0.462 |
| 0.4089 | 0.3498 | 5.319 | –1.754 | 1.0000 | 1.0000 | 15.689 | |
| 0.5014 | 0.4389 | 6.314 | –1.939 | | | | |
| | | pentan-1-ol (1) + TEA (2) ; *T*/K = 298.15 | | | | | |
| 0.0000 | 0.0000 | 2.417 | | 0.5968 | 0.5347 | 7.316 | –1.888 |
| 0.0588 | 0.0463 | 2.716 | –0.289 | 0.7072 | 0.6522 | 8.943 | –1.752 |
| 0.1062 | 0.0844 | 2.969 | –0.519 | 0.7960 | 0.7518 | 10.533 | –1.427 |
| 0.1482 | 0.1190 | 3.212 | –0.715 | 0.8517 | 0.8168 | 11.652 | –1.133 |
| 0.2159 | 0.1761 | 3.644 | –1.008 | 0.8939 | 0.8674 | 12.576 | –0.851 |
| 0.3011 | 0.2506 | 4.276 | –1.322 | 0.9465 | 0.9321 | 13.822 | –0.426 |
| 0.4089 | 0.3494 | 5.202 | –1.650 | 1.0000 | 1.0000 | 15.110 | |
| 0.5014 | 0.4384 | 6.162 | –1.820 | | | | |
| | | pentan-1-ol (1) + TEA (2) ; *T*/K = 303.15 | | | | | |
| 0.0000 | 0.0000 | 2.397 | | 0.5968 | 0.5342 | 7.123 | –1.759 |
| 0.0588 | 0.0462 | 2.685 | –0.273 | 0.7072 | 0.6517 | 8.683 | –1.626 |
| 0.1062 | 0.0843 | 2.928 | –0.492 | 0.7960 | 0.7514 | 10.201 | –1.318 |
| 0.1482 | 0.1188 | 3.164 | –0.675 | 0.8517 | 0.8165 | 11.265 | –1.044 |
| 0.2159 | 0.1758 | 3.583 | –0.948 | 0.8939 | 0.8672 | 12.144 | –0.781 |
| 0.3011 | 0.2503 | 4.193 | –1.243 | 0.9465 | 0.9320 | 13.320 | –0.391 |
| 0.4089 | 0.3490 | 5.087 | –1.547 | 1.0000 | 1.0000 | 14.537 | |
| 0.5014 | 0.4379 | 6.010 | –1.703 | | | | |
| | | heptan-1-ol (1) + TEA (2) ; *T*/K = 293.15 | | | | | |
| 0.0000 | 0.0000 | 2.438 | | 0.5950 | 0.5987 | 6.560 | –1.606 |
| 0.0451 | 0.0458 | 2.663 | –0.213 | 0.6948 | 0.6981 | 7.608 | –1.509 |
| 0.1029 | 0.1043 | 2.973 | –0.463 | 0.7925 | 0.7950 | 8.806 | –1.238 |
| 0.1466 | 0.1486 | 3.212 | –0.648 | 0.8478 | 0.8498 | 9.584 | –0.984 |
| 0.1997 | 0.2022 | 3.527 | –0.845 | 0.8982 | 0.8996 | 10.352 | –0.692 |
| 0.2979 | 0.3012 | 4.163 | –1.157 | 0.9465 | 0.9473 | 11.135 | –0.366 |
| 0.3963 | 0.4000 | 4.855 | –1.410 | 1.0000 | 1.0000 | 12.005 | |
| 0.4950 | 0.4989 | 5.647 | –1.564 | | | | |
| | | heptan-1-ol (1) + TEA (2) ; *T*/K = 298.15 | | | | | |
| 0.0000 | 0.0000 | 2.418 | | 0.5950 | 0.5982 | 6.398 | –1.455 |
| 0.0451 | 0.0457 | 2.635 | –0.198 | 0.6948 | 0.6976 | 7.398 | –1.358 |
| 0.1029 | 0.1041 | 2.933 | –0.431 | 0.7925 | 0.7947 | 8.532 | –1.107 |
| 0.1466 | 0.1483 | 3.167 | –0.598 | 0.8478 | 0.8495 | 9.268 | –0.869 |
| 0.1997 | 0.2019 | 3.470 | –0.782 | 0.8982 | 0.8994 | 9.979 | –0.611 |
| 0.2979 | 0.3007 | 4.086 | –1.064 | 0.9465 | 0.9472 | 10.706 | –0.318 |
| 0.3963 | 0.3995 | 4.752 | –1.296 | 1.0000 | 1.0000 | 11.504 | |
| 0.4950 | 0.4984 | 5.518 | –1.428 | | | | |
| | | heptan-1-ol (1) + TEA (2) ; *T*/K = 303.15 | | | | | |
| 0.0000 | 0.0000 | 2.397 | | 0.5950 | 0.5977 | 6.233 | –1.314 |
| 0.0451 | 0.0456 | 2.607 | –0.183 | 0.6948 | 0.6972 | 7.187 | –1.217 |
| 0.1029 | 0.1039 | 2.896 | –0.396 | 0.7925 | 0.7944 | 8.262 | –0.980 |
| 0.1466 | 0.1480 | 3.122 | –0.550 | 0.8478 | 0.8493 | 8.954 | –0.761 |
| 0.1997 | 0.2015 | 3.414 | –0.719 | 0.8982 | 0.8992 | 9.611 | –0.534 |
| 0.2979 | 0.3003 | 4.010 | –0.974 | 0.9465 | 0.9471 | 10.282 | –0.275 |
| 0.3963 | 0.3990 | 4.652 | –1.183 | 1.0000 | 1.0000 | 11.013 | |
| 0.4950 | 0.4978 | 5.389 | –1.297 | | | | |

[a] The standard uncertainties are: $u(T)$ = 0.02 K; $u(p)$ = 1 kPa; $u(\nu)$= 20 Hz; $u(x_1)$ = 0.0010; $u(\phi_1)$ = 0.004. The relative standard uncertainty is: $u_r(\varepsilon_r)$ = 0.003; and the relative combined expanded uncertainty (0.95 level of confidence) is $U_{rc}(\varepsilon_r^E) = 0.03$.

Table 4

Volume fractions of alkan-1-ol, $\phi_1$, refractive indices at the sodium D-line, $n_D$, and excess refractive indices at the sodium D-line, $n_D^E$, of alkan-1-ol (1) + TEA (2) liquid mixtures as functions of the mole fraction of the alkan-1-ol, $x_1$, at temperature $T$ and pressure $p$ = 0.1 MPa. [a]

| $x_1$ | $\phi_1$ | $n_D$ | $10^5 n_D^E$ | $x_1$ | $\phi_1$ | $n_D$ | $10^5 n_D^E$ |
|---|---|---|---|---|---|---|---|
| | | | methanol (1) + TEA (2) ; $T$/K = 293.15 | | | | |
| 0.0000 | 0.0000 | 1.40044 | | 0.6002 | 0.3040 | 1.38753 | 852 |
| 0.0588 | 0.0179 | 1.40013 | 94 | 0.6953 | 0.3990 | 1.38081 | 857 |
| 0.1072 | 0.0338 | 1.39980 | 173 | 0.7943 | 0.5291 | 1.37056 | 764 |
| 0.1765 | 0.0587 | 1.39918 | 285 | 0.8504 | 0.6232 | 1.36254 | 641 |
| 0.2036 | 0.0692 | 1.39890 | 331 | 0.9003 | 0.7243 | 1.35360 | 479 |
| 0.2989 | 0.1104 | 1.39750 | 481 | 0.9510 | 0.8496 | 1.34215 | 247 |
| 0.4053 | 0.1655 | 1.39510 | 629 | 0.9851 | 0.9506 | 1.33303 | 76 |
| 0.5193 | 0.2391 | 1.39143 | 782 | 1.0000 | 1.0000 | 1.32863 | |
| | | | methanol (1) + TEA (2) ; $T$/K = 298.15 | | | | |
| 0.0000 | 0.0000 | 1.39775 | | 0.6002 | 0.3039 | 1.38506 | 858 |
| 0.0588 | 0.0178 | 1.39752 | 101 | 0.6953 | 0.3989 | 1.37844 | 867 |
| 0.1072 | 0.0337 | 1.39719 | 178 | 0.7943 | 0.5290 | 1.36825 | 773 |
| 0.1765 | 0.0587 | 1.39661 | 294 | 0.8504 | 0.6231 | 1.36028 | 649 |
| 0.2036 | 0.0692 | 1.39635 | 341 | 0.9003 | 0.7242 | 1.35146 | 494 |
| 0.2989 | 0.1103 | 1.39493 | 486 | 0.9510 | 0.8495 | 1.34008 | 262 |
| 0.4053 | 0.1654 | 1.39271 | 649 | 0.9851 | 0.9506 | 1.33092 | 82 |
| 0.5193 | 0.2391 | 1.38891 | 786 | 1.0000 | 1.0000 | 1.32649 | |
| | | | methanol (1) + TEA (2) ; $T$/K = 303.15 | | | | |
| 0.0000 | 0.0000 | 1.39503 | | 0.6002 | 0.3038 | 1.38258 | 864 |
| 0.0588 | 0.0178 | 1.39482 | 102 | 0.6953 | 0.3988 | 1.37601 | 873 |
| 0.1072 | 0.0337 | 1.39452 | 181 | 0.7943 | 0.5289 | 1.36595 | 784 |
| 0.1765 | 0.0586 | 1.39396 | 297 | 0.8504 | 0.6230 | 1.35795 | 652 |
| 0.2036 | 0.0692 | 1.39371 | 346 | 0.9003 | 0.7241 | 1.34908 | 486 |
| 0.2989 | 0.1103 | 1.39236 | 495 | 0.9510 | 0.8494 | 1.33782 | 259 |
| 0.4053 | 0.1654 | 1.39003 | 644 | 0.9851 | 0.9505 | 1.32875 | 81 |
| 0.5193 | 0.2390 | 1.38631 | 784 | 1.0000 | 1.0000 | 1.32435 | |
| | | | propan-1-ol (1) + TEA (2) ; $T$/K = 293.15 | | | | |
| 0.0000 | 0.0000 | 1.40044 | | 0.6011 | 0.4475 | 1.40027 | 667 |
| 0.0476 | 0.0262 | 1.40088 | 84 | 0.6989 | 0.5551 | 1.39836 | 641 |
| 0.0989 | 0.0557 | 1.40121 | 162 | 0.7959 | 0.6770 | 1.39548 | 540 |
| 0.1620 | 0.0941 | 1.40154 | 254 | 0.8455 | 0.7463 | 1.39359 | 457 |
| 0.1967 | 0.1163 | 1.40171 | 304 | 0.9017 | 0.8314 | 1.39090 | 319 |
| 0.2936 | 0.1826 | 1.40195 | 430 | 0.9484 | 0.9081 | 1.38836 | 183 |
| 0.4005 | 0.2642 | 1.40203 | 562 | 1.0000 | 1.0000 | 1.38511 | |
| 0.4973 | 0.3471 | 1.40153 | 639 | | | | |
| | | | propan-1-ol (1) + TEA (2) ; $T$/K = 298.15 | | | | |
| 0.0000 | 0.0000 | 1.39775 | | 0.6011 | 0.4472 | 1.39795 | 675 |
| 0.0476 | 0.0261 | 1.39822 | 85 | 0.6989 | 0.5547 | 1.39615 | 653 |
| 0.0989 | 0.0556 | 1.39864 | 170 | 0.7959 | 0.6767 | 1.39338 | 555 |
| 0.1620 | 0.0940 | 1.39904 | 266 | 0.8455 | 0.7460 | 1.39142 | 461 |
| 0.1967 | 0.1162 | 1.39921 | 316 | 0.9017 | 0.8312 | 1.38887 | 332 |
| 0.2936 | 0.1824 | 1.39954 | 446 | 0.9484 | 0.9080 | 1.38629 | 187 |
| 0.4005 | 0.2639 | 1.39949 | 560 | 1.0000 | 1.0000 | 1.38306 | |
| 0.4973 | 0.3468 | 1.39902 | 635 | | | | |
| propan-1-ol (1) + TEA (2) ; $T$/K = 303.15 | | | | | | | |
| 0.0000 | 0.0000 | 1.39503 | | 0.6011 | 0.4468 | 1.39556 | 679 |
| 0.0476 | 0.0261 | 1.39553 | 86 | 0.6989 | 0.5544 | 1.39378 | 652 |
| 0.0989 | 0.0556 | 1.39595 | 170 | 0.7959 | 0.6764 | 1.39103 | 548 |
| 0.1620 | 0.0939 | 1.39642 | 270 | 0.8455 | 0.7458 | 1.38916 | 459 |
| 0.1967 | 0.1160 | 1.39665 | 324 | 0.9017 | 0.8310 | 1.38667 | 330 |
| 0.2936 | 0.1822 | 1.39700 | 452 | 0.9484 | 0.9078 | 1.38419 | 190 |
| 0.4005 | 0.2637 | 1.39705 | 571 | 1.0000 | 1.0000 | 1.38099 | |
| 0.4973 | 0.3465 | 1.39662 | 644 | | | | |
| butan-1-ol (1) + TEA (2) ; $T$/K = 293.15 | | | | | | | |
| 0.0000 | 0.0000 | 1.40044 | | 0.6063 | 0.5033 | 1.40628 | 642 |
| 0.0556 | 0.0373 | 1.40139 | 99 | 0.7054 | 0.6117 | 1.40572 | 598 |
| 0.1033 | 0.0705 | 1.40215 | 179 | 0.8013 | 0.7263 | 1.40448 | 488 |
| 0.1482 | 0.1027 | 1.40284 | 252 | 0.8498 | 0.7882 | 1.40353 | 400 |
| 0.1992 | 0.1406 | 1.40356 | 328 | 0.8978 | 0.8525 | 1.40241 | 295 |
| 0.2974 | 0.2178 | 1.40473 | 454 | 0.9530 | 0.9303 | 1.40085 | 148 |
| 0.3982 | 0.3033 | 1.40566 | 557 | 1.0000 | 1.0000 | 1.39929 | |
| 0.5027 | 0.3994 | 1.40624 | 626 | | | | |
| butan-1-ol (1) + TEA (2) ; $T$/K = 298.15 | | | | | | | |
| 0.0000 | 0.0000 | 1.39775 | | 0.6063 | 0.5029 | 1.40397 | 645 |
| 0.0556 | 0.0372 | 1.39877 | 104 | 0.7054 | 0.6113 | 1.40352 | 605 |
| 0.1033 | 0.0703 | 1.39958 | 186 | 0.8013 | 0.7259 | 1.40234 | 492 |
| 0.1482 | 0.1026 | 1.40029 | 259 | 0.8498 | 0.7880 | 1.40148 | 408 |
| 0.1992 | 0.1404 | 1.40107 | 338 | 0.8978 | 0.8523 | 1.40036 | 299 |
| 0.2974 | 0.2175 | 1.40228 | 463 | 0.9530 | 0.9302 | 1.39884 | 151 |
| 0.3982 | 0.3029 | 1.40324 | 563 | 1.0000 | 1.0000 | 1.39730 | |
| 0.5027 | 0.3990 | 1.40385 | 628 | | | | |
| butan-1-ol (1) + TEA (2) ; $T$/K = 303.15 | | | | | | | |
| 0.0000 | 0.0000 | 1.39503 | | 0.6063 | 0.5024 | 1.40168 | 652 |
| 0.0556 | 0.0372 | 1.39616 | 112 | 0.7054 | 0.6109 | 1.40125 | 606 |
| 0.1033 | 0.0702 | 1.39701 | 196 | 0.8013 | 0.7256 | 1.40015 | 493 |
| 0.1482 | 0.1024 | 1.39776 | 270 | 0.8498 | 0.7877 | 1.39932 | 409 |
| 0.1992 | 0.1402 | 1.39852 | 345 | 0.8978 | 0.8521 | 1.39827 | 302 |
| 0.2974 | 0.2172 | 1.39982 | 473 | 0.9530 | 0.9300 | 1.39680 | 153 |
| 0.3982 | 0.3026 | 1.40078 | 567 | 1.0000 | 1.0000 | 1.39529 | |
| 0.5027 | 0.3986 | 1.40150 | 637 | | | | |
| pentan-1-ol (1) + TEA (2) ; $T$/K = 293.15 | | | | | | | |
| 0.0000 | 0.0000 | 1.40044 | | 0.5966 | 0.5349 | 1.41188 | 636 |
| 0.0473 | 0.0372 | 1.40173 | 94 | 0.6985 | 0.6431 | 1.41225 | 570 |
| 0.1017 | 0.0809 | 1.40321 | 200 | 0.8018 | 0.7588 | 1.41214 | 449 |
| 0.1483 | 0.1193 | 1.40432 | 274 | 0.8501 | 0.8152 | 1.41197 | 379 |
| 0.2080 | 0.1696 | 1.40566 | 361 | 0.8985 | 0.8732 | 1.41156 | 283 |
| 0.2989 | 0.2490 | 1.40765 | 484 | 0.9499 | 0.9365 | 1.41079 | 146 |
| 0.4002 | 0.3417 | 1.40941 | 572 | 1.0000 | 1.0000 | 1.40993 | |
| 0.4963 | 0.4339 | 1.41092 | 635 | | | | |
| | | pentan-1-ol (1) + TEA (2) ; $T$/K = 298.15 | | | | | |
| 0.0000 | 0.0000 | 1.39775 | | 0.5966 | 0.5345 | 1.40966 | 645 |
| 0.0473 | 0.0371 | 1.39912 | 99 | 0.6985 | 0.6427 | 1.41021 | 590 |
| 0.1017 | 0.0808 | 1.40057 | 199 | 0.8018 | 0.7585 | 1.41012 | 463 |
| 0.1483 | 0.1191 | 1.40176 | 279 | 0.8501 | 0.8149 | 1.40986 | 380 |
| 0.2080 | 0.1693 | 1.40332 | 384 | 0.8985 | 0.8730 | 1.40944 | 279 |
| 0.2989 | 0.2487 | 1.40516 | 487 | 0.9499 | 0.9364 | 1.40873 | 144 |
| 0.4002 | 0.3412 | 1.40707 | 583 | 1.0000 | 1.0000 | 1.40794 | |
| 0.4963 | 0.4334 | 1.40855 | 637 | | | | |
| | | pentan-1-ol (1) + TEA (2) ; $T$/K = 303.15 | | | | | |
| 0.0000 | 0.0000 | 1.39503 | | 0.5966 | 0.5340 | 1.40737 | 651 |
| 0.0473 | 0.0370 | 1.39653 | 110 | 0.6985 | 0.6422 | 1.40796 | 593 |
| 0.1017 | 0.0806 | 1.39806 | 215 | 0.8018 | 0.7581 | 1.40796 | 467 |
| 0.1483 | 0.1189 | 1.39927 | 294 | 0.8501 | 0.8146 | 1.40777 | 386 |
| 0.2080 | 0.1691 | 1.40089 | 401 | 0.8985 | 0.8728 | 1.40733 | 279 |
| 0.2989 | 0.2483 | 1.40272 | 498 | 0.9499 | 0.9363 | 1.40669 | 146 |
| 0.4002 | 0.3408 | 1.40473 | 598 | 1.0000 | 1.0000 | 1.40592 | |
| 0.4963 | 0.4329 | 1.40620 | 645 | | | | |
| | | heptan-1-ol (1) + TEA (2) ; $T$/K = 293.15 | | | | | |
| 0.0000 | 0.0000 | 1.40044 | | 0.5968 | 0.6005 | 1.42052 | 569 |
| 0.0521 | 0.0529 | 1.40283 | 112 | 0.6945 | 0.6978 | 1.42217 | 502 |
| 0.0980 | 0.0994 | 1.40484 | 201 | 0.7957 | 0.7982 | 1.42340 | 386 |
| 0.1540 | 0.1560 | 1.40719 | 300 | 0.8465 | 0.8485 | 1.42384 | 310 |
| 0.2017 | 0.2042 | 1.40913 | 378 | 0.8935 | 0.8950 | 1.42413 | 229 |
| 0.2992 | 0.3025 | 1.41266 | 495 | 0.9479 | 0.9487 | 1.42431 | 120 |
| 0.3958 | 0.3995 | 1.41568 | 565 | 1.0000 | 1.0000 | 1.42433 | |
| 0.4951 | 0.4990 | 1.41832 | 591 | | | | |
| | | heptan-1-ol (1) + TEA (2) ; $T$/K = 298.15 | | | | | |
| 0.0000 | 0.0000 | 1.39775 | | 0.5968 | 0.6000 | 1.41831 | 574 |
| 0.0521 | 0.0528 | 1.40021 | 115 | 0.6945 | 0.6973 | 1.42004 | 508 |
| 0.0980 | 0.0992 | 1.40230 | 209 | 0.7957 | 0.7979 | 1.42135 | 393 |
| 0.1540 | 0.1558 | 1.40469 | 308 | 0.8465 | 0.8482 | 1.42183 | 318 |
| 0.2017 | 0.2039 | 1.40664 | 384 | 0.8935 | 0.8948 | 1.42213 | 234 |
| 0.2992 | 0.3020 | 1.41026 | 503 | 0.9479 | 0.9486 | 1.42234 | 123 |
| 0.3958 | 0.3990 | 1.41334 | 572 | 1.0000 | 1.0000 | 1.42236 | |
| 0.4951 | 0.4985 | 1.41607 | 600 | | | | |
| | | heptan-1-ol (1) + TEA (2) ; $T$/K = 303.15 | | | | | |
| 0.0000 | 0.0000 | 1.39503 | | 0.5968 | 0.5995 | 1.41611 | 581 |
| 0.0521 | 0.0527 | 1.39757 | 119 | 0.6945 | 0.6969 | 1.41794 | 517 |
| 0.0980 | 0.0990 | 1.39971 | 215 | 0.7957 | 0.7975 | 1.41930 | 399 |
| 0.1540 | 0.1555 | 1.40218 | 317 | 0.8465 | 0.8480 | 1.41980 | 322 |
| 0.2017 | 0.2035 | 1.40415 | 392 | 0.8935 | 0.8946 | 1.42014 | 238 |
| 0.2992 | 0.3016 | 1.40783 | 510 | 0.9479 | 0.9485 | 1.42036 | 125 |
| 0.3958 | 0.3985 | 1.41100 | 580 | 1.0000 | 1.0000 | 1.42041 | |
| 0.4951 | 0.4979 | 1.41379 | 607 | | | | |

[a] The standard uncertainties are: $u(T)$ = 0.02 K; $u(p)$ = 1 kPa; $u(x_1)$ = 0.0010; $u(\phi_1)$ = 0.004, $u(n_D)$ = 0.00008. The combined expanded uncertainty (0.95 level of confidence) is $U_c(n_D^E)$ = 0.0002.

Table 5

Coefficients $A_i$ and standard deviations, $\sigma\left(F^E\right)$ (equation (6)), for the representation of $F^E$ at temperature $T$ and pressure $p$ = 0.1 MPa for alkan-1-ol (1) + TEA liquid mixtures by equation (5).

| Property $F^E$ | alkan-1-ol | $T$/K | $A_0$ | $A_1$ | $A_2$ | $A_3$ | $A_3$ | $\sigma\left(F^E\right)$ |
|---|---|---|---|---|---|---|---|---|
| $\varepsilon_r^E$ | methanol | 293.15 | –3.50 | 1.8 | 4.4 | 7.0 | 5.7 | 0.02 |
| | | 298.15 | –3.43 | 1.8 | 4.2 | 6.6 | 5.5 | 0.02 |
| | | 303.15 | –3.37 | 1.7 | 4.0 | 6.4 | 5.4 | 0.02 |
| | propan-1-ol | 293.15 | –7.71 | –2.7 | 1.4 | 2.0 | | 0.018 |
| | | 298.15 | –7.39 | –2.5 | 1.5 | 2.0 | | 0.017 |
| | | 303.15 | –7.06 | –2.4 | 1.5 | 1.9 | | 0.014 |
| | butan-1-ol | 293.15 | –8.09 | –3.4 | 0.4 | 1.6 | | 0.015 |
| | | 298.15 | –7.67 | –3.1 | 0.5 | 1.6 | | 0.014 |
| | | 303.15 | –7.26 | –2.9 | 0.6 | 1.5 | | 0.014 |
| | pentan-1-ol | 293.15 | –7.81 | –2.70 | | | | 0.019 |
| | | 298.15 | –7.31 | –2.43 | | | | 0.018 |
| | | 303.15 | –6.82 | –2.15 | | | | 0.018 |
| | heptan-1-ol | 293.15 | –6.33 | –1.80 | | | | 0.018 |
| | | 298.15 | –5.75 | –1.46 | | | | 0.017 |
| | | 303.15 | –5.19 | –1.15 | | | | 0.018 |
| $10^5 n_D^E$ | methanol | 293.15 | 3050 | 2224 | 809 | | | 8 |
| | | 298.15 | 3080 | 2236 | 918 | | | 5 |
| | | 303.15 | 3090 | 2233 | 945 | | | 8 |
| | propan-1-ol | 293.15 | 2547 | 1161 | 251 | | | 4 |
| | | 298.15 | 2557 | 1164 | 412 | | | 3 |
| | | 303.15 | 2587 | 1120 | 356 | | | 1.4 |
| | butan-1-ol | 293.15 | 2500 | 832 | 136 | | | 2 |
| | | 298.15 | 2515 | 816 | 221 | | | 3 |
| | | 303.15 | 2538 | 774 | 269 | | | 3 |
| | pentan-1-ol | 293.15 | 2528 | 551 | | | | 7 |
| | | 298.15 | 2573 | 545 | | | | 6 |
| | | 303.15 | 2585 | 475 | 205 | | | 6 |
| | heptan-1-ol | 293.15 | 2366 | –16 | –30 | 144 | | 1.1 |
| | | 298.15 | 2392 | –23 | 28 | 148 | | 1.1 |
| | | 303.15 | 2422 | –16 | 69 | 101 | | 0.9 |
| $\left(\partial\varepsilon_r^E/\partial T\right)_p$ / $K^{-1}$ | methanol | 298.15 | 0.014 | –0.014 | –0.04 | –0.06 | –0.03 | 0.0005 |
| | propan-1-ol | 298.15 | 0.0661 | 0.030 | | | | 0.0005 |
| | butan-1-ol | 298.15 | 0.0832 | 0.047 | 0.014 | | | 0.0004 |
| | pentan-1-ol | 298.15 | 0.0953 | 0.056 | 0.017 | | | 0.0004 |
| | heptan-1-ol | 298.15 | 0.1079 | 0.065 | 0.029 | | | 0.0004 |

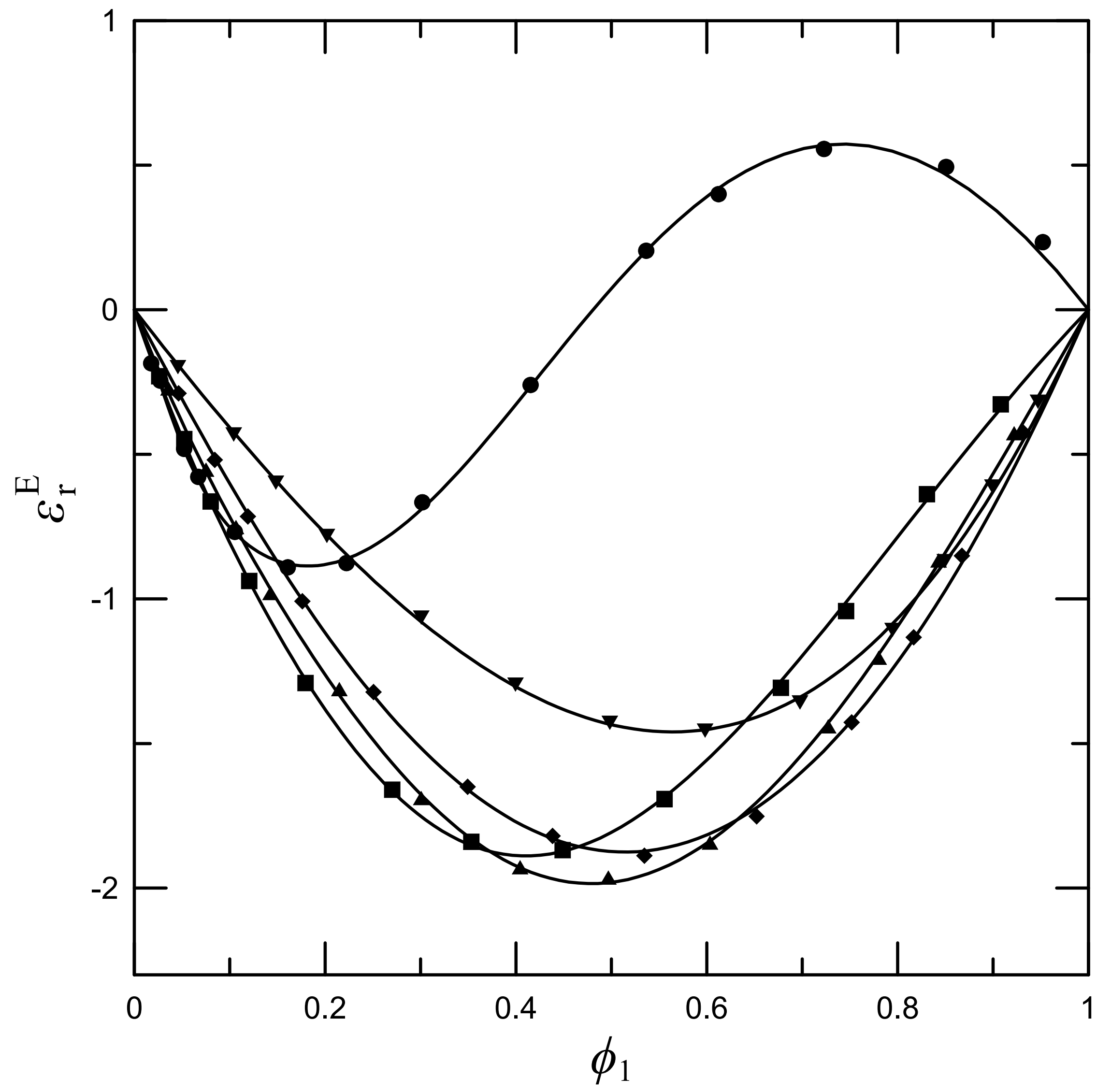

Figure 1

Excess relative permittivity, $\varepsilon_r^E$, of alkan-1-ol (1) + TEA (2) liquid mixtures as a function of the alkan-1-ol volume fraction, $\phi_1$, at 0.1 MPa, 298.15 K and 1 MHz. Full symbols, experimental values (this work): (●), methanol; (■), propan-1-ol; (▲), butan-1-ol; (♦), pentan-1-ol; (▼), heptan-1-ol. Solid lines, calculations with equation (5) using the coefficients from Table 5.

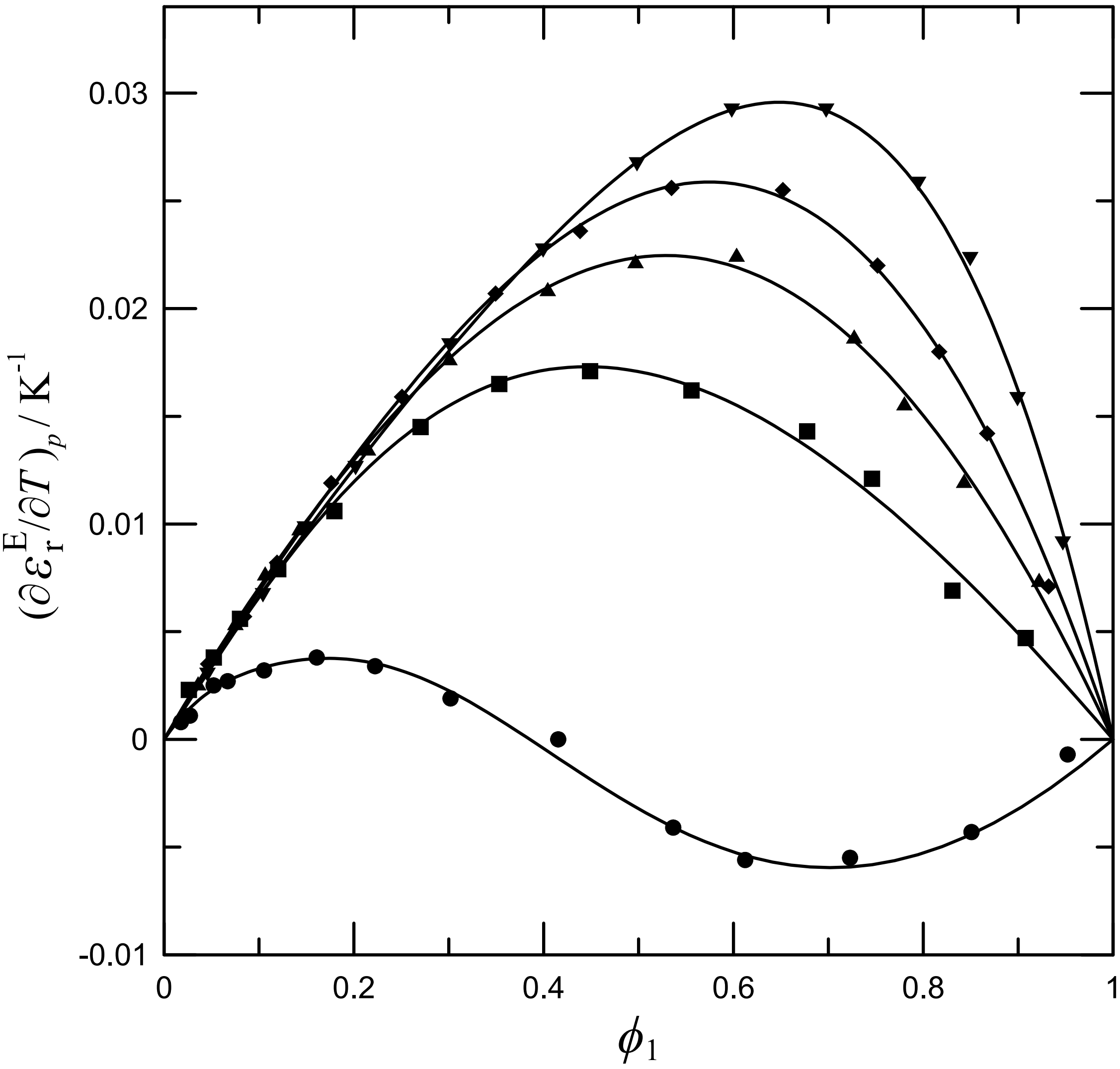

Figure 2

Temperature derivative of the excess relative permittivity, $\left(\partial \varepsilon_r^E / \partial T\right)_p$, of alkan-1-ol (1) + TEA (2) liquid mixtures as a function of the alkan-1-ol volume fraction, $\phi_1$, at 0.1 MPa, 298.15 K and 1 MHz. Full symbols, experimental values (this work): (●), methanol; (■), propan-1-ol; (▲), butan-1-ol; (♦), pentan-1-ol; (▼), heptan-1-ol. Solid lines, calculations with equation (5) using the coefficients from Table 5.

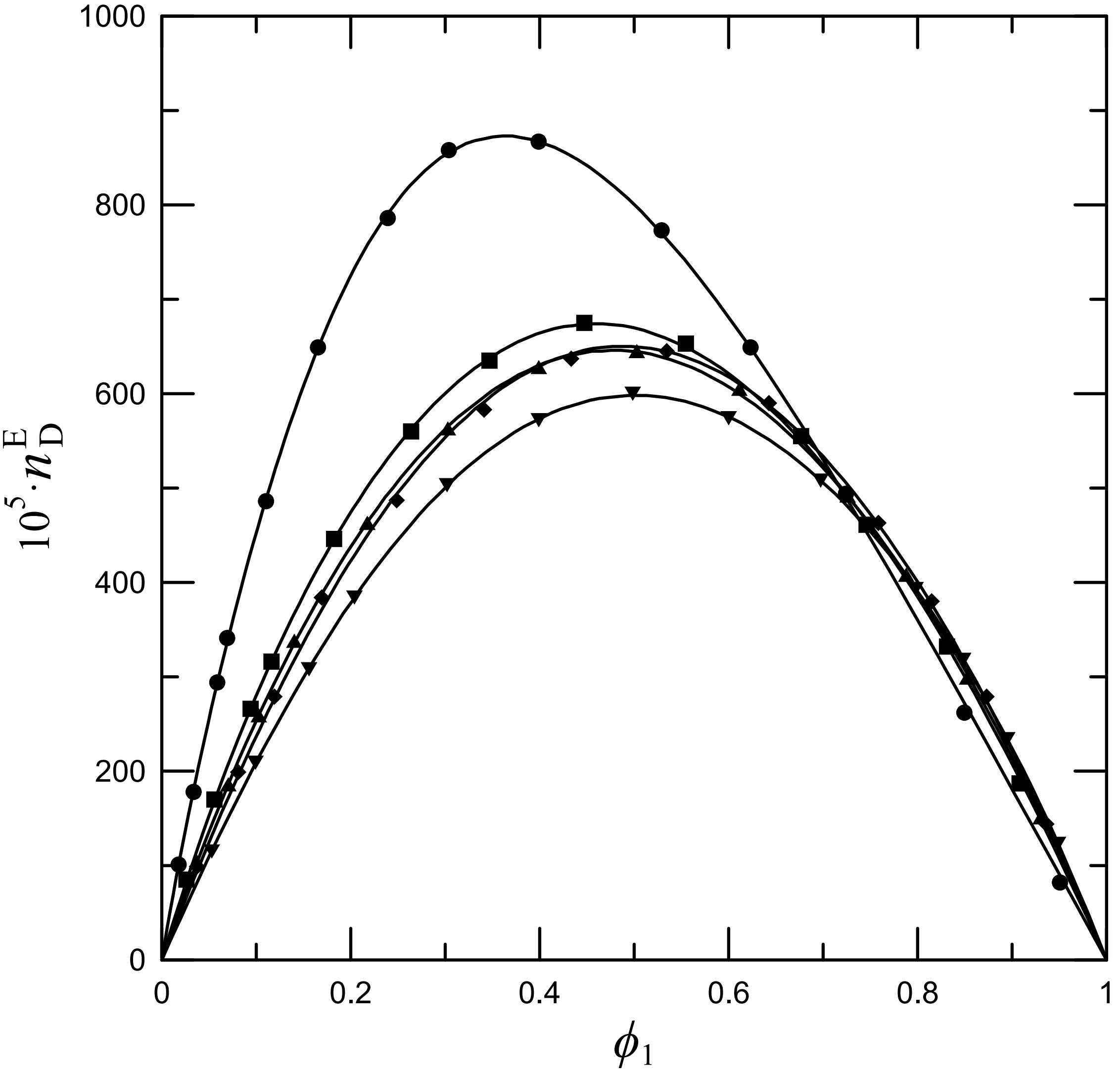

Figure 3

Excess refractive index at the sodium D-line, $n_D^E$, of alkan-1-ol (1) + TEA (2) liquid mixtures as a function of the alkan-1-ol volume fraction, $\phi_1$, at 0.1 MPa, 298.15 K. Full symbols, experimental values (this work): (●), methanol; (■), propan-1-ol; (▲), butan-1-ol; (♦), pentan-1-ol; (▼), heptan-1-ol. Solid lines, calculations with equation (5) using the coefficients from Table 5.

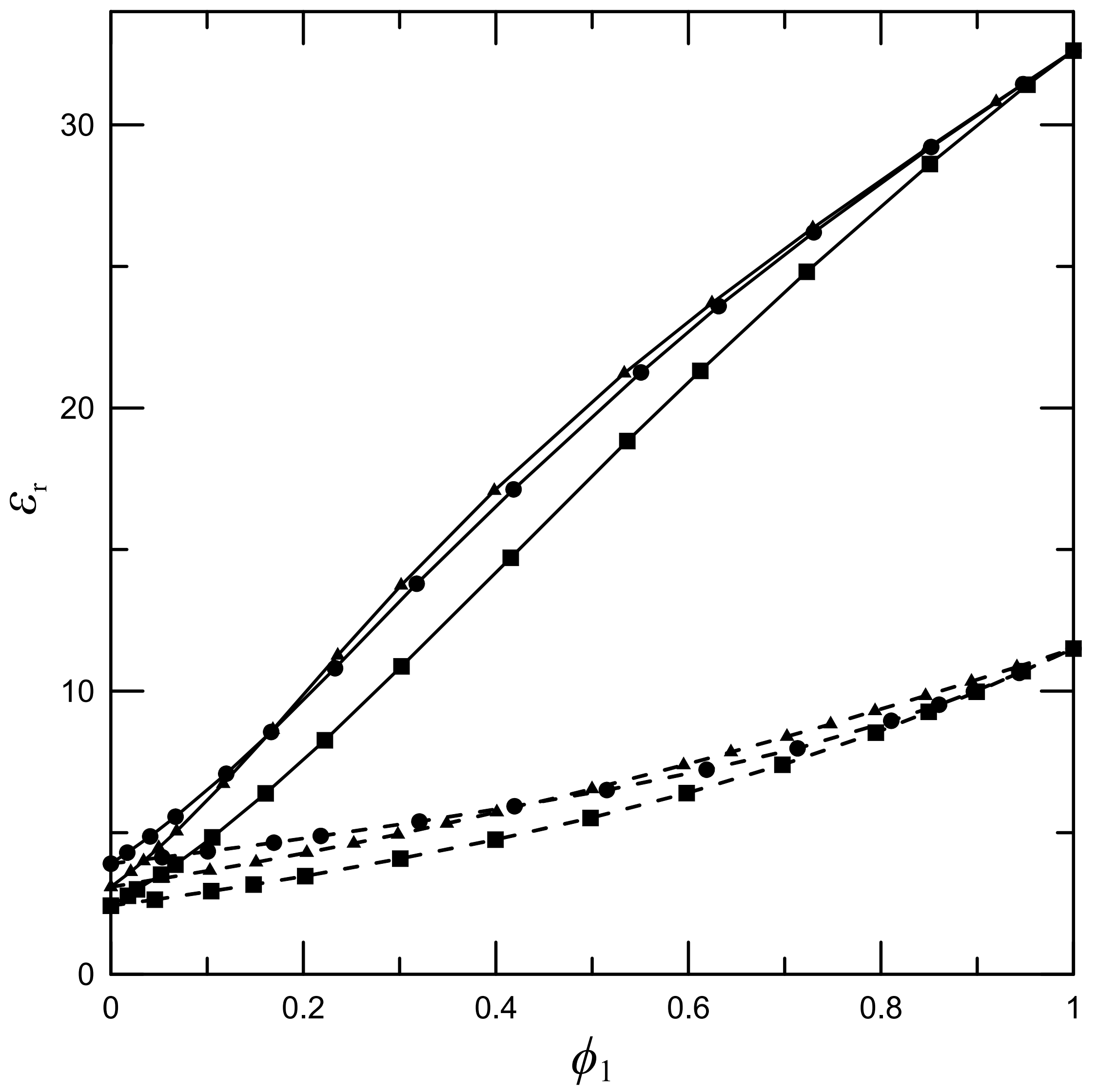

Figure 4

Relative permittivity, $\varepsilon_r$, of alkan-1-ol (1) + amine (2) liquid mixtures as a function of the alkan-1-ol volume fraction, $\phi_1$, at 0.1 MPa, 298.15 K and 1 MHz: (●), HxA [27]; (▲), DPA [28]; (■), TEA (this work). Solid lines, methanol; dashed lines, heptan-1-ol.

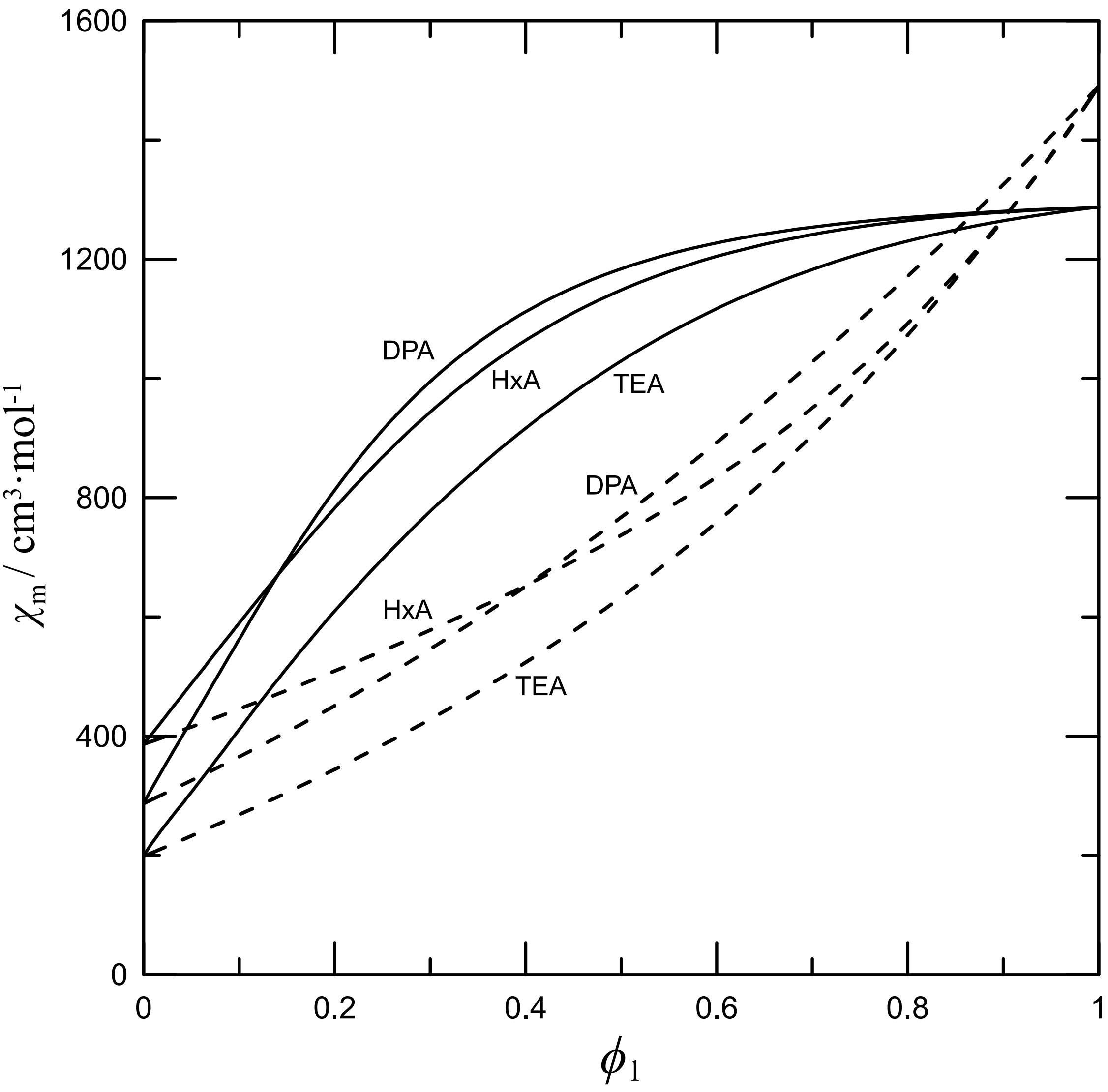

Figure 5

Molar susceptibility, $\chi_m$, of alkan-1-ol (1) + amine (2) liquid mixtures as a function of the alkan-1-ol volume fraction, $\phi_1$, at 0.1 MPa, 298.15 K and 1 MHz: HxA [27]; DPA [28]; TEA (this work). Solid lines, methanol; dashed lines, heptan-1-ol.

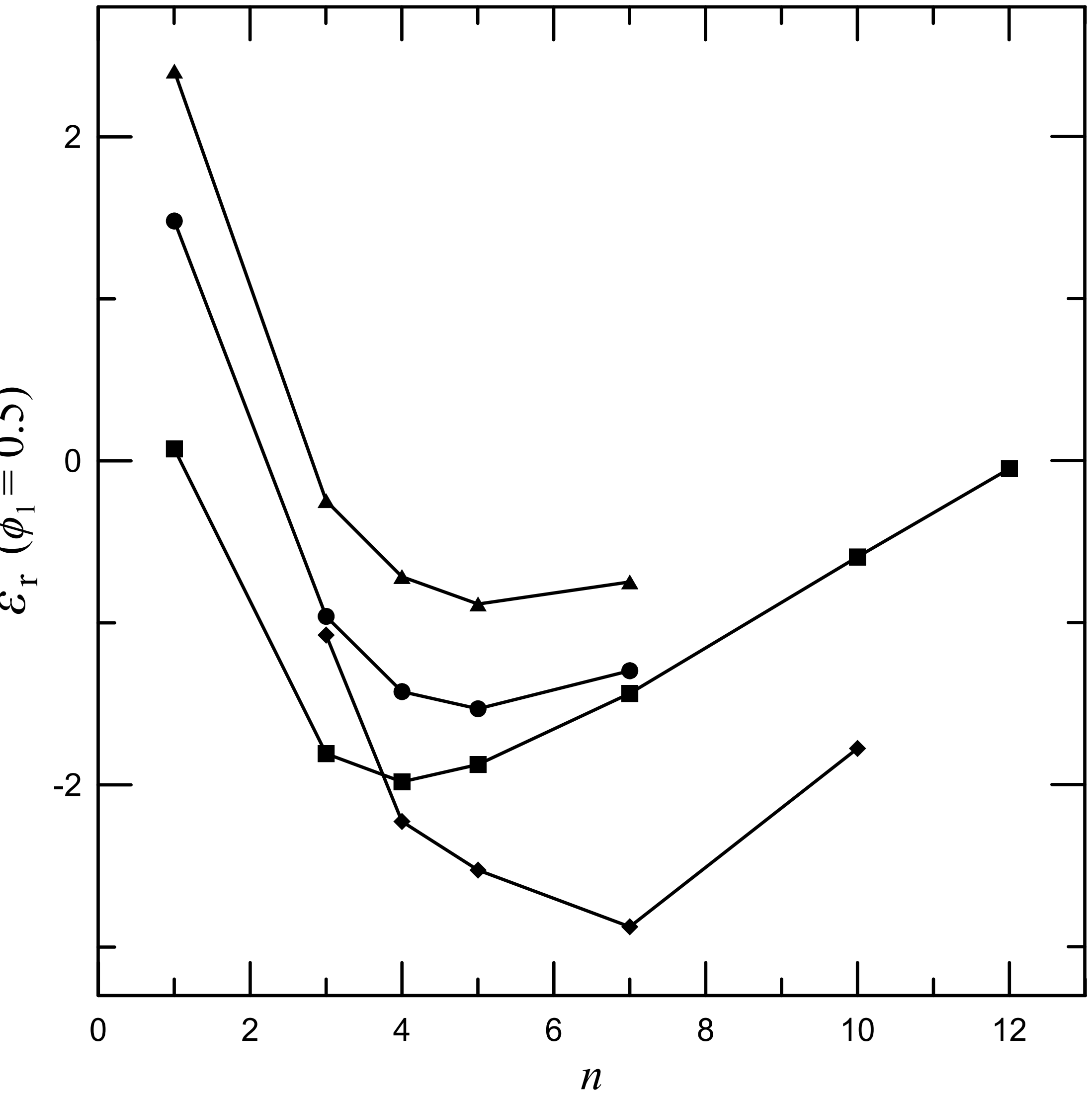

Figure 6

Excess relative permittivity at $\phi_1 = 0.5$ ($\phi_1$, alkan-1-ol volume fraction) of alkan-1-ol (1) + amine (2) or + heptane (2) liquid mixtures as a function of the number of carbon atoms of the alkan-1-ol, *n*, at 0.1 MPa, 298.15 K and 1 MHz: (●), HxA [27]; (▲), DPA [28]; (■), TEA (*n* = 1 to 7, this work; *n* = 10,12 are literature values [55] at 293.15 K); (♦), heptane [24,52,53].

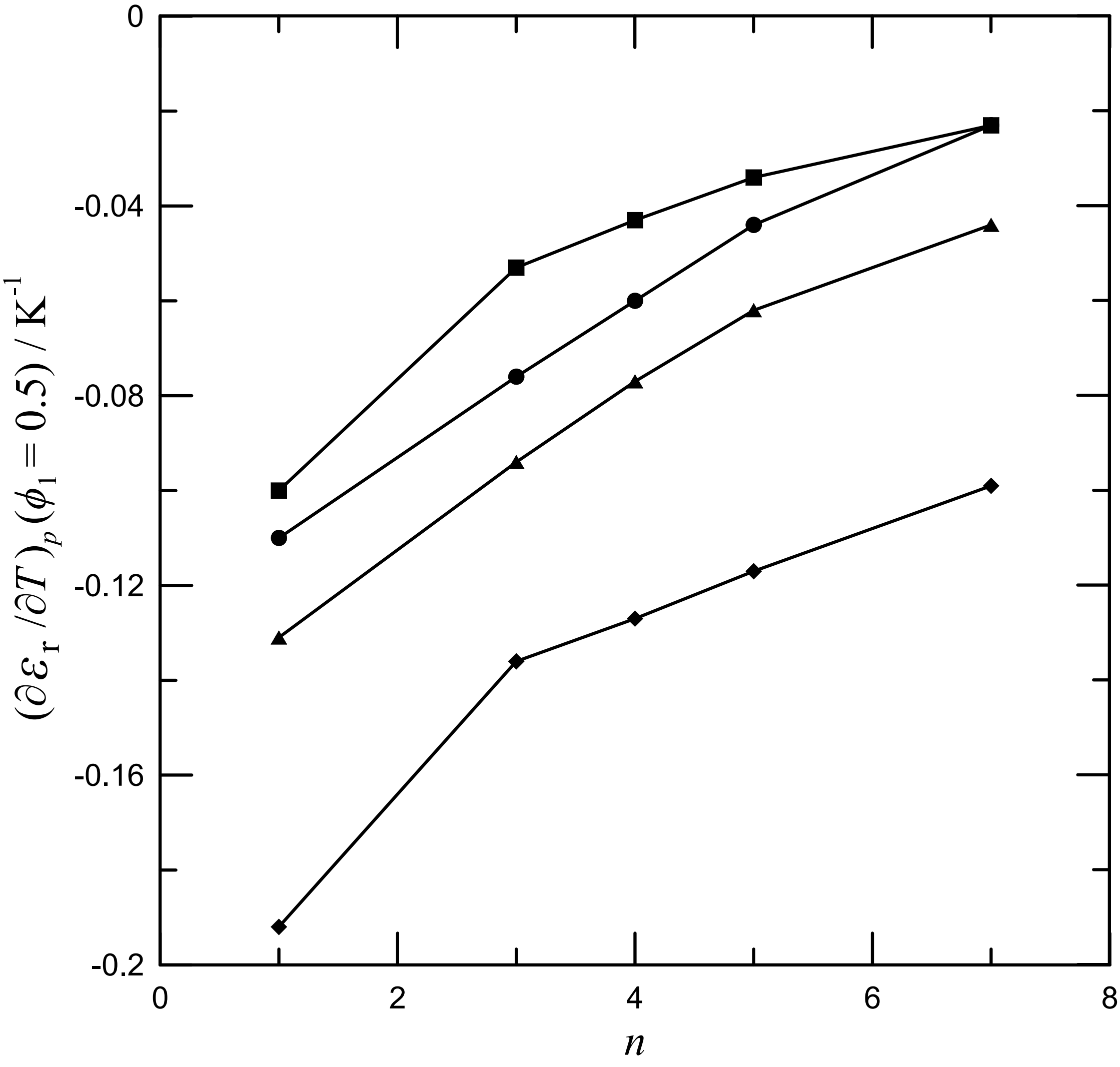

Figure 7

Temperature derivative of the relative permittivity, $(\partial\varepsilon_r / \partial T)_p$, at $\phi_1 = 0.5$ ($\phi_1$, alkan-1-ol volume fraction) of alkan-1-ol (1) + amine (2) liquid mixtures or pure alkan-1-ols as a function of the number of carbon atoms of the alkan-1-ol, *n*, at 0.1 MPa, 298.15 K and 1 MHz: (●), HxA [27]; (▲), DPA [28]; (■), TEA (this work); (♦), pure alkan-1-ols (this work).

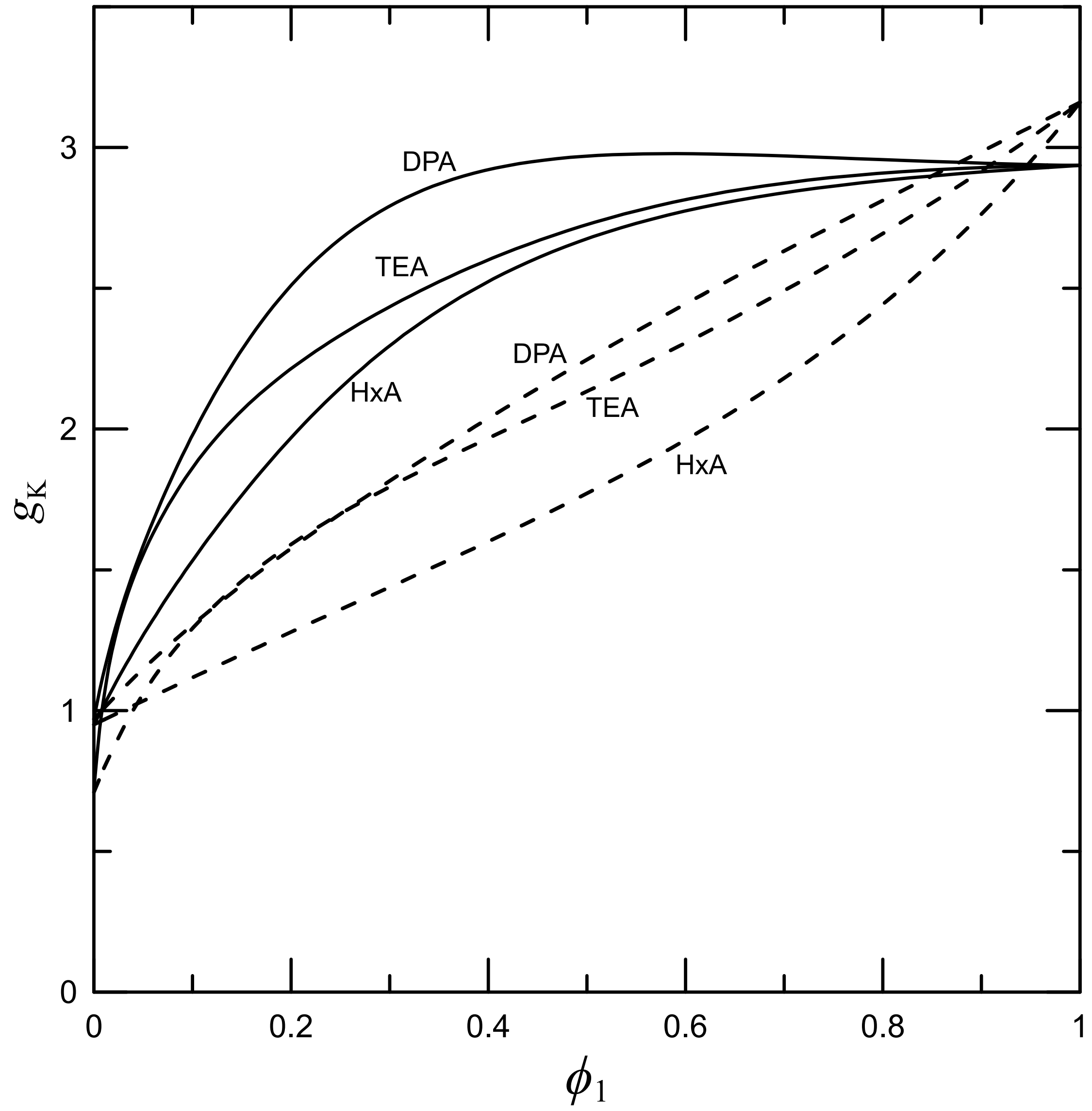

Figure 8

Kirkwood correlation factor, $g_K$, of alkan-1-ol (1) +amine (2) liquid mixtures as a function of the alkan-1-ol volume fraction, $\phi_1$, at 0.1 MPa, 298.15 K: HxA [27]; DPA [28]; TEA (this work). Solid lines, methanol; dashed lines, heptan-1-ol.

# Thermodynamics of mixtures with strongly negative deviations from Raoult's law. XVII. Permittivities and refractive indices for alkan-1-ol + *N,N*-diethylethanamine systems at (293.15-303.15) K. Application of the Kirkwood-Fröhlich model

## Supplementary material

Fernando Hevia, Juan Antonio González*, Ana Cobos, Isaías García de la Fuente, L.F. Sanz

G.E.T.E.F., Departamento de Física Aplicada, Facultad de Ciencias, Universidad de Valladolid. Paseo de Belén, 7, 47011 Valladolid, Spain.

*e-mail: jagl@termo.uva.es; Tel: +34 983 423757

Table S1

Volume fraction of alkan-1-ol, $\phi_1$, and derivative of the excess relative permittivity at frequency $\nu$ = 1 MHz, $\left(\partial\varepsilon_r^E/\partial T\right)_p$, of alkan-1-ol (1) + *N,N*-diethylethanamine (TEA) (2) liquid mixtures as functions of the mole fraction of the alkan-1-ol, $x_1$, at temperature *T* and pressure *p* = 0.1 MPa. [a]

| $x_1$ | $\phi_1$ | $\left(\partial\varepsilon_r^E/\partial T\right)_p$ / K$^{-1}$ | $x_1$ | $\phi_1$ | $\left(\partial\varepsilon_r^E/\partial T\right)_p$ / K$^{-1}$ |
|---|---|---|---|---|---|
| | | methanol (1) + TEA (2) ; *T*/K = 298.15 | | | |
| 0.0580 | 0.0176 | 0.0008 | 0.5979 | 0.3019 | 0.0019 |
| 0.0878 | 0.0272 | 0.0011 | 0.7096 | 0.4154 | 0.0000 |
| 0.1593 | 0.0522 | 0.0025 | 0.7992 | 0.5365 | –0.0041 |
| 0.1981 | 0.0670 | 0.0027 | 0.8445 | 0.6123 | –0.0056 |
| 0.2880 | 0.1053 | 0.0032 | 0.8997 | 0.7229 | –0.0055 |
| 0.3971 | 0.1608 | 0.0038 | 0.9515 | 0.8509 | –0.0043 |
| 0.4959 | 0.2224 | 0.0034 | 0.9856 | 0.9522 | –0.0007 |
| | | propan-1-ol (1) + TEA (2) ; *T*/K = 298.15 | | | |
| 0.0476 | 0.0261 | 0.0023 | 0.6027 | 0.4488 | 0.0171 |
| 0.0932 | 0.0523 | 0.0038 | 0.6997 | 0.5557 | 0.0162 |
| 0.1394 | 0.0800 | 0.0056 | 0.7966 | 0.6776 | 0.0143 |
| 0.2032 | 0.1204 | 0.0079 | 0.8455 | 0.7460 | 0.0121 |
| 0.2896 | 0.1795 | 0.0106 | 0.9014 | 0.8307 | 0.0069 |
| 0.4084 | 0.2704 | 0.0145 | 0.9484 | 0.9080 | 0.0047 |
| 0.5043 | 0.3532 | 0.0165 | | | |
| | | butan-1-ol (1) + TEA (2) ; *T*/K = 298.15 | | | |
| 0.0536 | 0.0359 | 0.0026 | 0.6006 | 0.4969 | 0.0222 |
| 0.1102 | 0.0752 | 0.0054 | 0.6985 | 0.6034 | 0.0225 |
| 0.1536 | 0.1065 | 0.0077 | 0.8023 | 0.7272 | 0.0187 |
| 0.2025 | 0.1429 | 0.0098 | 0.8438 | 0.7801 | 0.0156 |
| 0.2941 | 0.2148 | 0.0135 | 0.8910 | 0.8430 | 0.0120 |
| 0.3957 | 0.3007 | 0.0177 | 0.9475 | 0.9222 | 0.0074 |
| 0.5082 | 0.4043 | 0.0209 | | | |
| | | pentan-1-ol (1) + TEA (2) ; *T*/K = 298.15 | | | |
| 0.0588 | 0.0463 | 0.0035 | 0.5968 | 0.5347 | 0.0256 |
| 0.1062 | 0.0844 | 0.0057 | 0.7072 | 0.6522 | 0.0255 |
| 0.1482 | 0.1190 | 0.0082 | 0.7960 | 0.7518 | 0.0220 |
| 0.2159 | 0.1761 | 0.0119 | 0.8517 | 0.8168 | 0.0180 |
| 0.3011 | 0.2506 | 0.0159 | 0.8939 | 0.8674 | 0.0142 |
| 0.4089 | 0.3494 | 0.0207 | 0.9465 | 0.9321 | 0.0071 |
| 0.5014 | 0.4384 | 0.0236 | | | |
| heptan-1-ol (1) + TEA (2) ; $T$/K = 298.15 | | | | | |
| 0.0451 | 0.0457 | 0.0030 | 0.5950 | 0.5982 | 0.0292 |
| 0.1029 | 0.1041 | 0.0067 | 0.6948 | 0.6976 | 0.0292 |
| 0.1466 | 0.1483 | 0.0098 | 0.7925 | 0.7947 | 0.0258 |
| 0.1997 | 0.2019 | 0.0126 | 0.8478 | 0.8495 | 0.0223 |
| 0.2979 | 0.3007 | 0.0183 | 0.8982 | 0.8994 | 0.0158 |
| 0.3963 | 0.3995 | 0.0227 | 0.9465 | 0.9472 | 0.0091 |
| 0.4950 | 0.4984 | 0.0267 | | | |

[a] The standard uncertainties are: $u(T)$ = 0.02 K; $u(p)$ = 1 kPa; $u(\nu)$ = 20 Hz; $u(x_1)$ = 0.0010; $u(\phi_1)$ = 0.004; $u\left[\left(\partial\varepsilon_r^E/\partial T\right)_p\right]$ = 0.0008 K$^{-1}$.

Table S2

Values of the derivative of permittivity with respect to temperature for pure compounds, $\left(\partial\varepsilon_{\mathrm{r}}^{*}/\partial T\right)_{p}$, and for mixtures, $\left(\partial\varepsilon_{\mathrm{r}}/\partial T\right)_{p}$, at $\phi_1$ = 0.5 ($\phi_1$, volume fraction of the alkan-1-ol), temperature $T$ = 298.15 K and pressure $p$ = 0.1 MPa. [a]

| Compound | $\left(\partial\varepsilon_{\mathrm{r}}^{*}/\partial T\right)_{p}$/K$^{-1}$ | | $\left(\partial\varepsilon_{\mathrm{r}}/\partial T\right)_{p}$/K$^{-1}$ | | |
|---|---|---|---|---|---|
| | Exp. | Lit. | alkan-1-ol + HxA [1] | alkan-1-ol + DPA [2] | alkan-1-ol + TEA |
| Methanol | –0.192 | –0.195 [3] | –0.110 | –0.131 | –0.100 |
| propan-1-ol | –0.136 | –0.130 [4] | –0.076 | –0.094 | –0.053 |
| butan-1-ol | –0.127 | –0.122 [4] | –0.060 | –0.077 | –0.043 |
| pentan-1-ol | –0.117 | –0.110 [4] | –0.044 | –0.062 | –0.034 |
| heptan-1-ol | –0.099 | –0.096 [4] | –0.023 | –0.044 | –0.023 |

[a] hexan-1-amine (HxA), *N*-propylpropan-1-amine (DPA), *N*,*N*-diethylethanamine (TEA).

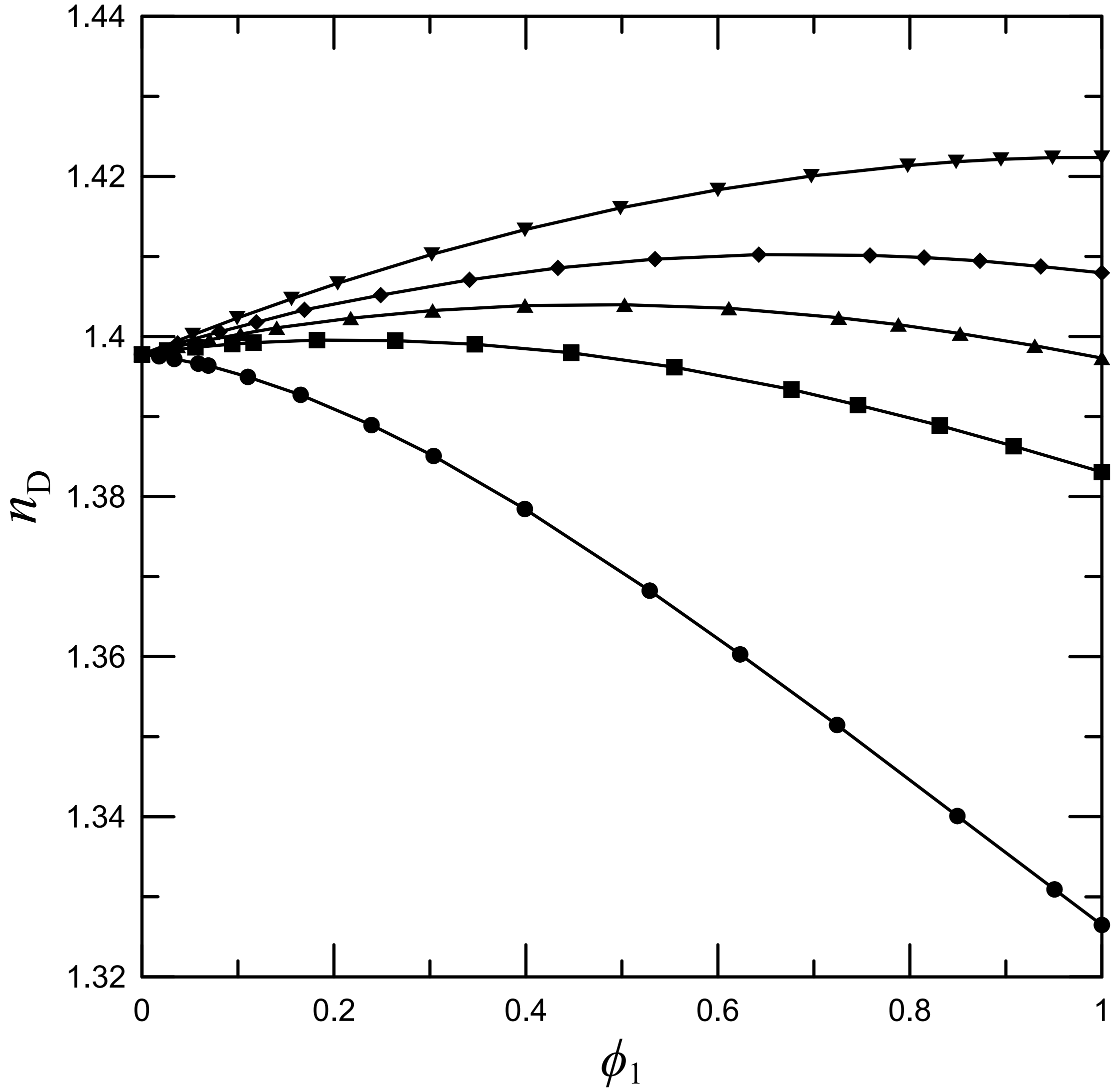

Figure S1

Refractive index at the sodium D-line, $n_D$, of alkan-1-ol (1) + TEA (2) liquid mixtures as a function of the alkan-1-ol volume fraction, $\phi_1$, at 0.1 MPa, 298.15 K. Full symbols, experimental values (this work): (●), methanol; (■), propan-1-ol; (▲), butan-1-ol; (♦), pentan-1-ol; (▼), heptan-1-ol.

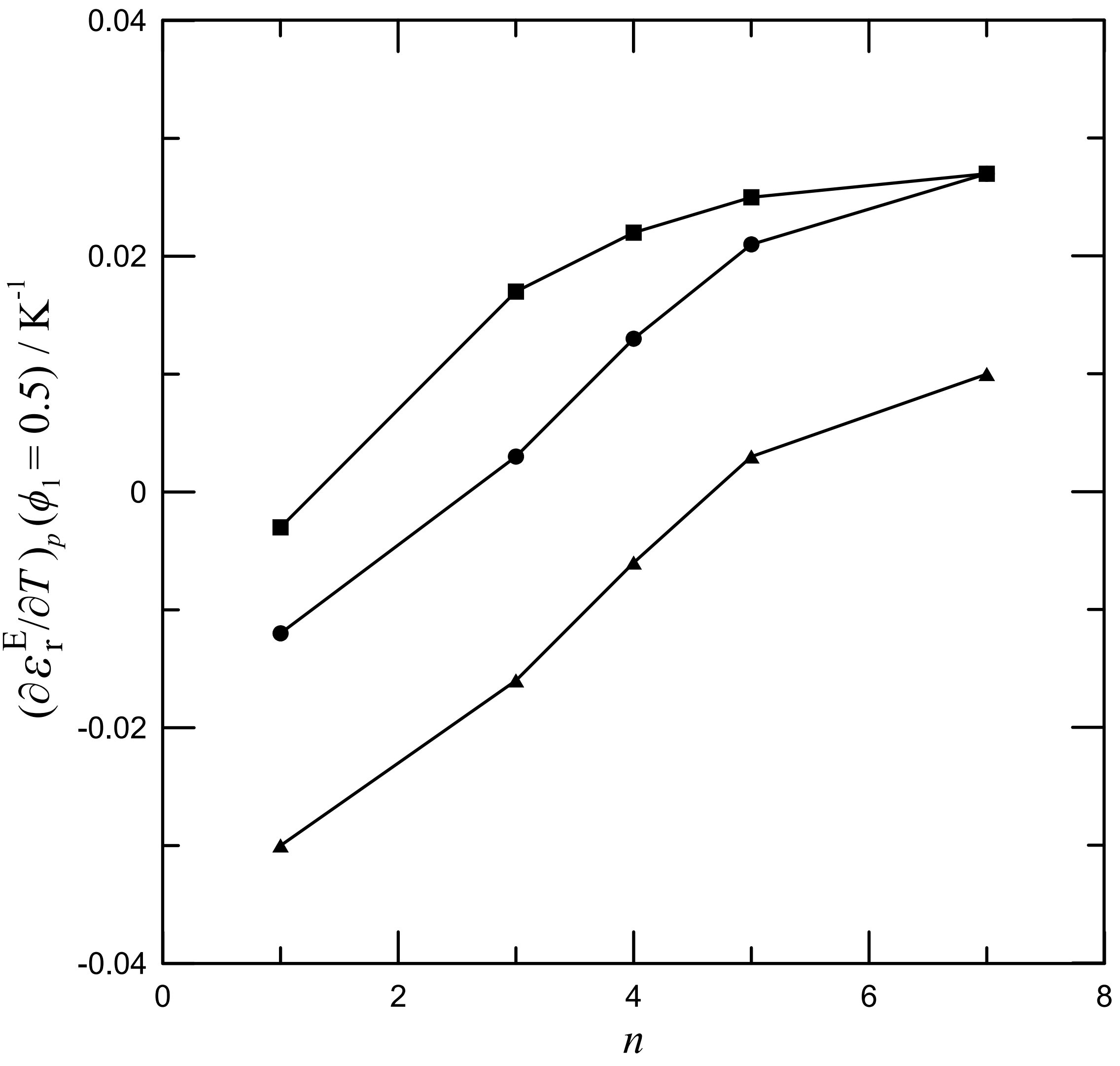

Figure S2

Temperature derivative of the excess relative permittivity, $\left(\partial\varepsilon_r^E / \partial T\right)_p$, at $\phi_1 = 0.5$ of alkan-1-ol (1) + amine (2) liquid mixtures as a function of the number of carbon atoms of the alkan-1-ol, *n*, at 0.1 MPa, 298.15 K and 1 MHz: (●), HxA [1]; (▲), DPA [2]; (■), TEA (this work).

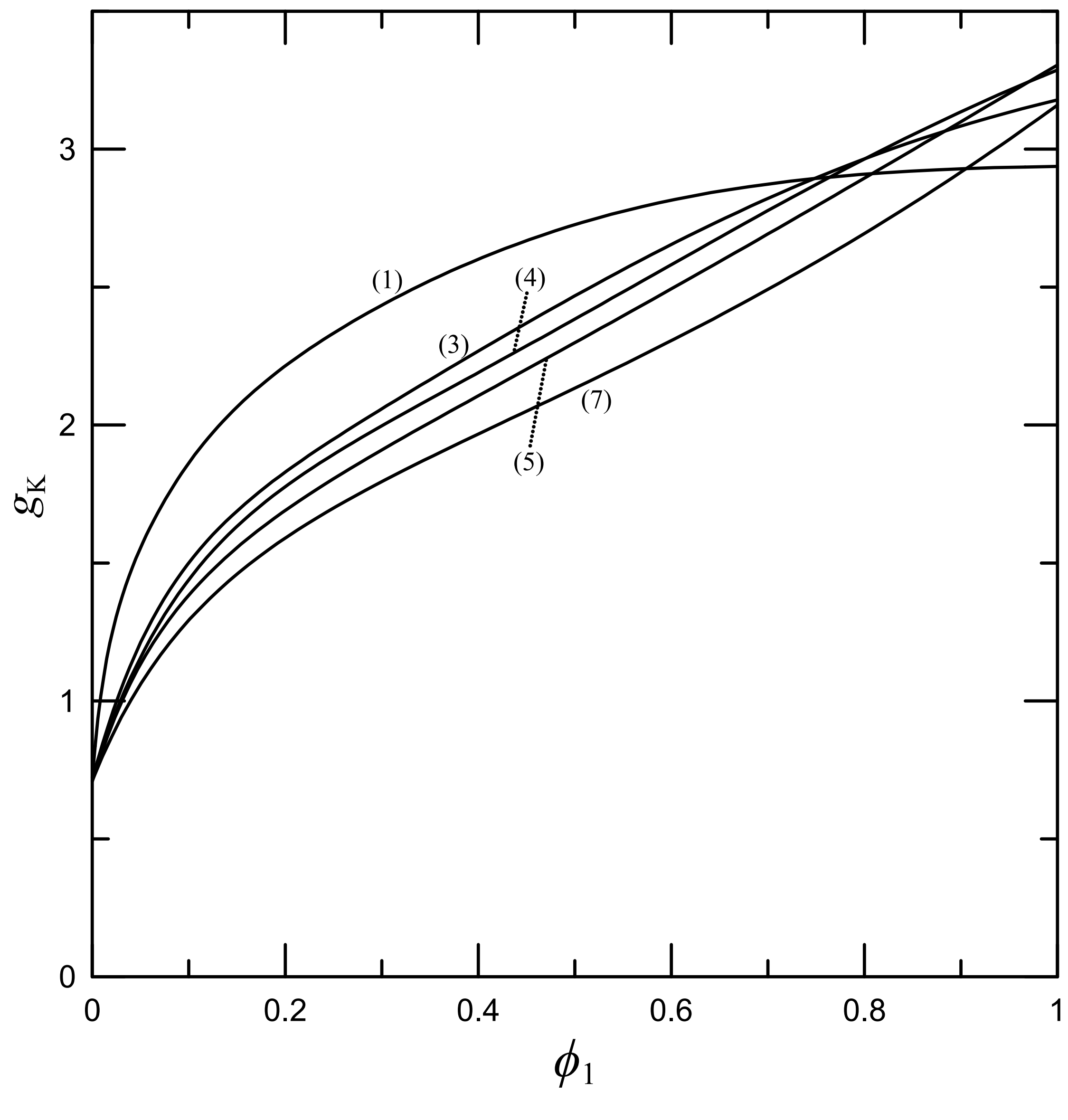

Figure S3

Kirkwood correlation factor, $g_K$, of alkan-1-ol (1) + TEA (2) liquid mixtures as a function of the alkan-1-ol volume fraction, $\phi_1$, at 0.1 MPa, 298.15 K. Numbers in parentheses indicate the number of carbon atoms of the alkan-1-ol.

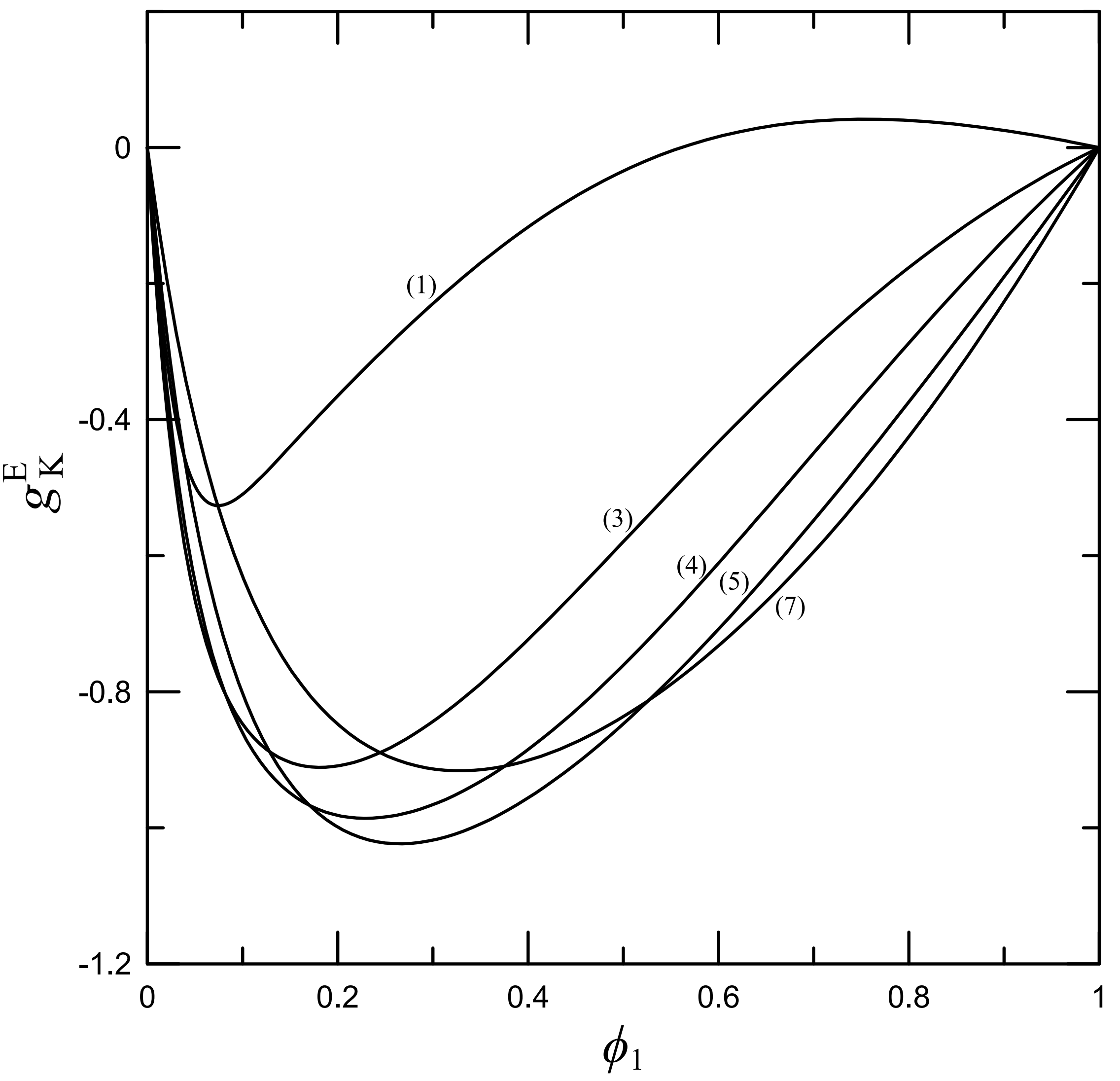

Figure S4

Excess Kirkwood correlation factor, $g_K^E$, of alkan-1-ol (1) + TEA (2) liquid mixtures as a function of the alkan-1-ol volume fraction, $\phi_1$, at 0.1 MPa, 298.15 K. Numbers in parentheses indicate the number of carbon atoms of the alkan-1-ol.

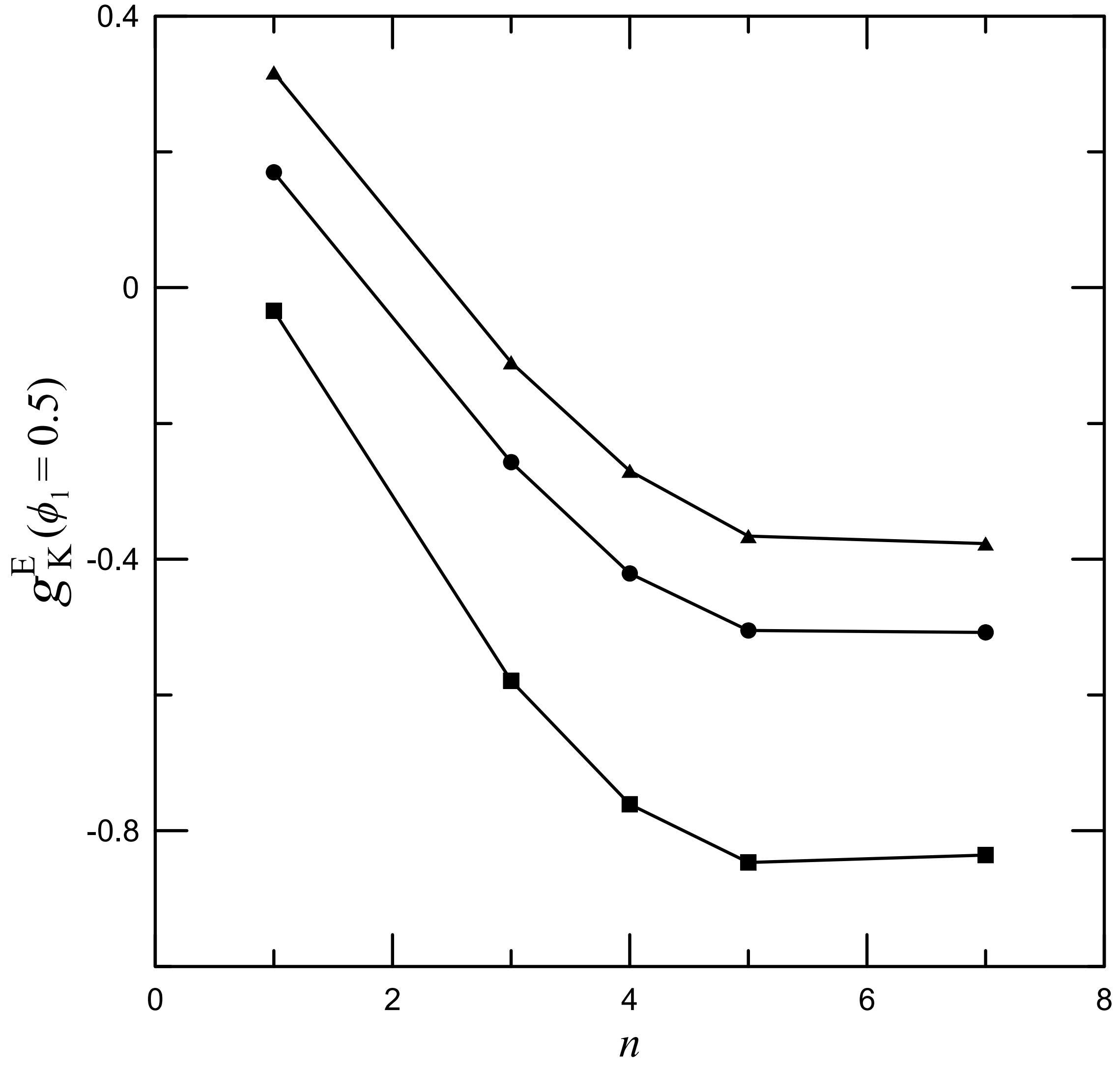

Figure S5

Excess Kirkwood correlation factor at $\phi_1 = 0.5$ ($\phi_1$, alkan-1-ol volume fraction) of alkan-1-ol (1) + amine (2) liquid mixtures as a function of the number of carbon atoms of the alkan-1-ol, at 0.1 MPa, 298.15 K: (●), HxA [1]; (▲), DPA [2]; (■), TEA (this work).

## References for supplementary material